\documentclass[a4paper,fleqn,usenatbib]{mnras}

\usepackage{newtxtext,newtxmath}

\usepackage[T1]{fontenc}
\usepackage{ae,aecompl}

\usepackage{graphicx}	
\usepackage{amsmath}	
\usepackage{amssymb}	

\usepackage{epstopdf}

\defcitealias{Robertson10}{R10}
\defcitealias{LopezSanjuan15}{LS15}

\title[Galaxy bias at redshift $z\sim2$ from cosmic variance]{Observational determination of the galaxy bias from cosmic variance with a random pointing survey: Clustering of $z \sim 2$ galaxies from Hubble's BoRG survey}

\author[A. J. Cameron et al.]{
Alex J. Cameron$^{1,2}$,\thanks{E-mail: alexc@student.unimelb.edu.au}
Michele Trenti$^{1,2}$,
Rachael C. Livermore$^{1,2}$,
\newauthor Cameron van der Velden$^{1}$
\\
$^{1}$The University of Melbourne, School of Physics, VIC 3010, Australia\\
$^{2}$ARC Centre of Excellence for All Sky Astrophysics in 3 Dimensions (ASTRO 3D)\\
}

\date{Accepted XXX. Received YYY; in original form ZZZ}

\pubyear{2018}

\begin{document}
\label{firstpage}
\pagerange{\pageref{firstpage}--\pageref{lastpage}}
\maketitle

\begin{abstract} 

Gravitational clustering broadens the count-in-cells distribution of galaxies for surveys along uncorrelated (well-separated) lines of sight beyond Poisson noise. A number of methods have proposed to measure this excess ``cosmic'' variance to constrain the galaxy bias (i.e. the strength of clustering) independently of the two-point correlation function. Here we present an observational application of these methods using data from 141 uncorrelated fields ($\sim700~\mathrm{arcmin^2}$ total) from Hubble's Brightest of Reionizing Galaxies (BoRG) survey. We use BoRG's broad-band imaging in optical and near infrared to identify $N\sim1000$ photometric candidates at $z\sim2$ through a combination of colour selection and photometric redshift determination, building a magnitude-limited sample with $m_{AB}\leq24.5$ in F160W. We detect a clear excess in the variance of the galaxy number counts distribution compared to Poisson expectations, from which we estimate a galaxy bias $b\approx3.63\pm0.57$. When divided by SED-fit classification into $\sim$400 early-type and $\sim$600 late-type candidates, we estimate biases of $b_\textit{early}\approx4.06\pm0.67$ and $b_\textit{late}\approx2.98\pm0.98$ respectively. These estimates are consistent with previous measurements of the bias from the two-point correlation function, and demonstrate that with $N\gtrsim100$ sight-lines, each containing $N\gtrsim5$ objects, the counts-in-cell analysis provides a robust measurement of the bias. This implies that the method can be applied effectively to determine clustering properties (and characteristic dark-matter halo masses) of $z\sim6-9$ galaxies from a pure-parallel \emph{James Webb Space Telescope} survey similar in design to Hubble's BoRG survey.

\end{abstract}

\begin{keywords}
galaxies: statistics -- methods: statistical -- surveys
\end{keywords}



\section{Introduction}

Understanding the connection between luminous galaxies and the dark matter haloes that host them presents an important challenge in the development of galaxy evolution models. This connection encodes the lasting effect of the complex array of astrophysical processes involved in the formation and evolution of galaxies, such as star formation, tidal stripping, and mergers, that are all connected to the dynamical assembly of their parent dark-matter haloes. While direct measurements of the dark-matter halo masses are limited to cases where dynamical tracers are available at large radii \citep[e.g.][]{Rood79, Kent&Gunn82, Millington86}, or through gravitational lensing \citep[e.g.][]{Tyson90, Dodelson04}, clustering measurements make it possible to draw statistical inference between the dark haloes and the properties of galaxies that inhabit them, through analytic and numerical studies of the dark-matter halo clustering \citep[e.g.][]{Mo&White96, Mo&White02}.

Observationally, galaxy clustering is typically quantified via the construction of a two-point correlation function ($\xi(r)$), describing the excess probability of finding a galaxy at separation $r$ from another galaxy. The galaxy bias, that is the clustering strength of galaxies relative to the matter field, relates the correlation function of galaxies ($\xi_\text{gg}$) to that of the dark matter field ($\xi_\text{DM}$) as a simple ratio, conventionally defined at a separation $r=8$ Mpc.

\begin{equation}
$$ b^{2} \equiv \xi_\text{gg} / \xi_\text{DM} $$ .
\label{eq:bias_def}
\end{equation}

Thus, the observed galaxy correlation functions can be related to those derived for the dark matter field, enabling constraints to be identified on the mass of the dark matter haloes hosting these galaxies \citep{Ouchi04, Adelberger05, Hayashi07} as well as how galaxies might be distributed within these haloes \citep{Zehavi05, Ouchi05, Lee06}. Previous measurements of correlation functions conducted with large surveys such as 2dFGRS \citep{Colless01} and SDSS \citep{York00} have been successful in linking the clustering of galaxies to a range of intrinsic properties such as luminosity, colour and morphology in the local Universe and at low redshift \citep{Norberg01, Norberg02, Budavari03, Madgwick03, Zehavi05, Zehavi11, Li06, Wang07}.

Similar measurements have been extended progressively to higher redshifts \citep[e.g.][]{Blanc08, Meneux09, McCracken10, delaTorre11, Wake11, Lin12, Mostek13, BaroneNugent14, Skibba14, Sato14, Coil17, Ishikawa17, Harikane18}. Still, one key challenge for observations during the epoch of reionization ($z>6$) is the limited area available with suitably deep infrared data, which makes legacy surveys potentially affected by systematic uncertainties due to large-scale clustering, or ``cosmic variance'' \citep{Trenti08}. 

The impact of cosmic variance on high-$z$ galaxy surveys can however be exploited to quantify the clustering properties of the galaxies, if observations span a large number of independent (uncorrelated) lines of sight, as has been innovatively proposed in a theoretical paper by \citet[\citetalias{Robertson10} hereafter]{Robertson10}. The \citetalias{Robertson10} method is based essentially on a counts-in-cells approach \citep[e.g.][]{Peebles80, Efstathiou90, Efstathiou95, Andreani94, Adelberger98, Viironen18},  applied to a set of uncorrelated fields. In this framework, the number counts distribution will be broader than Poisson, and once the Poisson variance is removed, the remaining dispersion in the counts can be attributed to the effect of clustering, enabling a measurement of the galaxy bias.

Recently, \citet[\citetalias{LopezSanjuan15} hereafter]{LopezSanjuan15} applied a similar observational measurement to counts from the ALHAMBRA survey. In that study, the galaxy bias was successfully derived for $z\sim 0.7$ galaxies from the counts in the 48 subfields of the survey in 8 independent regions of the sky. These authors were able to compare this to bias values obtained from conventional correlation function measurements of the same survey, demonstrating consistency between the two measurements. The more sophisticated analysis employed by \citetalias{LopezSanjuan15} involves fitting a log-normal distribution to the set of number counts. Removing the estimated Poisson shot noise from this distribution yields the intrinsic scatter from which the cosmic variance can be derived. While the work by \citetalias{LopezSanjuan15} demonstrates the feasibility of this type of analysis, the number of independent lines of sight is small.

Here, we address this limitation and use a large \emph{Hubble Space Telescope} (\emph{HST}) random pointing survey, the Brightest of Reionizing Galaxies (BoRG) survey \citep{Trenti11, Bradley12, Schmidt14, Calvi16} to assess the performance of the \citetalias{LopezSanjuan15} and \citetalias{Robertson10} bias measurement methods using over 140 independent lines of sight. BoRG has been designed to identify rare bright ($L>L_*$) galaxies at $z\gtrsim 7$ from broad-band imaging, minimizing the effects of cosmic variance on the galaxy luminosity function. While the number counts of these rare sources do not allow an application of the method (Poisson noise dominates when the average number of counts per field is below unity), the $VYJH$ filter set of BoRG also proves ideal for the identification of $z\sim2$ galaxies via the 4000\,\AA{} break. Thus, in this paper we apply these methods to the counts of photometrically selected $z\sim2$ galaxies from BoRG to estimate their bias, and investigate how it depends on galaxy properties such as spectral type to demonstrate the efficiency of the technique and compare its results against those from the literature based on the two point correlation function.  

The paper is structured as follows. In section \ref{data} we introduce the BoRG survey and give an overview of the data we have used including some of the technical aspects. Section \ref{methods} outlines the methods we have used in creating source catalogs and selecting $z\sim2$ candidates, including the subtle differences between the two BoRG sub-samples, as well as how the \citetalias{LopezSanjuan15} and \citetalias{Robertson10} methods are applied to our number counts to calculate the large-scale galaxy bias. Section \ref{results} outlines the results of our analysis. In sections \ref{borgz8_results} and \ref{borgz9_results} we obtain a preliminary bias values for the BoRG[z8] and BoRG[z9] sub-samples considered separately. Section \ref{combined_sample_results} then details our final bias values and how they compare to previous measurements. We summarise in Section \ref{conclusion}, and discuss applications to future observations of high-$z$ galaxies with the \emph{James Webb Space Telescope}.

In this paper we adopt the \citet{Planck16} cosmology: $\Omega_\Lambda = 0.692$, $\Omega_M = 0.308$, $\sigma_8 = 0.815$ and $H_0 = 67.8$ km s$^{-1}$ Mpc$^{-1}$. All magnitudes are quoted in the $AB$ magnitude system \citep{Oke&Gunn83}.

\section{Data}
\label{data}

The Brightest of Reionizing Galaxies (BoRG) survey \citep{Trenti11} is an optical to near-infrared multi-waveband pure-parallel survey conducted with Wide-Field Camera 3 (WFC3) on \emph{HST} with the goal of finding the most luminous galaxies at redshifts $z \gtrsim 7$ using the Lyman-break dropout technique \citep{Steidel96}. 

The survey consists of two generations spanning a number of \emph{HST} cycles, with the focus shifting from identification of Lyman-break dropout candidates at $z\sim8$ in the first generation, BoRG[z8] \citep{Trenti11}, to pushing this search to \emph{HST}'s detection limit at $z\sim9-10$ in the second generation, BoRG[z9] \citep{Calvi16}.

The BoRG[z8] data set consists of 71 independent pointings imaged with WFC3/IR in three different near-infrared filters (F098M, F125W and F160W) with additional imaging conducted with WFC3/UVIS in one optical filter (either F606W or F600LP). In this paper we use the 2014 (DR3) public release\footnote{\url{https://archive.stsci.edu/prepds/borg/}}.

Additionally, we use data collected with BoRG[z9], consisting of the initial 28 fields presented in \citet{Calvi16} as well as 46 new pointings imaged with WFC3/IR in four different near-infrared filters (F105W, F125W, F140W and F160W) with optical imaging conducted with WFC3/UVIS in the F350LP filter.

The design of pure-parallel surveys is limited by constraints imposed by the primary observations, with exposure times varying in each available opportunity. This leads to a non-uniform survey, with  5$\sigma$ limiting magnitudes (for point sources and aperture $r=0''.2$) ranging between $m_\text{AB} = 25.6 - 27.5$. Also, the data do not include standard dithering (not employed by spectroscopic primary programs). However, a combination of careful planning to take advantage of natural roll-angle dithering between different orbits and of Laplacian edge detection filtering allows us to effectively identify and remove artifacts, such as hot/cold pixels that have not been flagged by the calibration pipeline. Overall, the image quality is very close to prime dithered observations (see \citealt{Calvi16} for a comprehensive discussion). For more detailed discussions of the technical details of the BoRG survey, the reader is referred to \citet{Trenti11, Trenti12, Bradley12, Schmidt14, Calvi16}.

Two fields, borg\_0751+2917 and borg\_1209+4543, were serendipitously imaged in both the BoRG[z8] and BoRG[z9] data sets, due to the same primary target being revisited during multiple cycles. To avoid counting the same fields twice, only the BoRG[z8] pointings were considered for these fields in our galaxy bias analysis. However, these fields presented a unique opportunity to compare the photometry of the two separate filter sets considered in our analysis. This comparison is discussed in section \ref{field_compare}.

From the remaining sample, one field was discarded due to a problematic over-density of stars which was covering the majority of the area in the image, while one was discarded due to significant Galactic dust-reddening which could have affected the photometric redshift selection if small scale fluctuations in reddening had been present. This left us with a combined sample of 141 uncorrelated fields spanning $\sim$700 arcmin$^2$.

\section{Methods}
\label{methods}

In this analysis, we identify $z\sim2$ candidates with a $Y-H$ colour selection, capturing the prominent break at 4000\,\AA{} in rest-frame luminosity, and subsequent photometric redshift estimation. The importance this places on the $Y$ band photometry leads us to initially consider the BoRG[z8] and BoRG[z9] data sets separately as different Y filters are employed (F098M vs. F105W). This section introduces the analysis methods generally and describes specific application to the BoRG[z8] sample before outlining how the difference in $Y$ band filter is mitigated.

\begin{figure}
	\centering
	\includegraphics[width=\columnwidth]{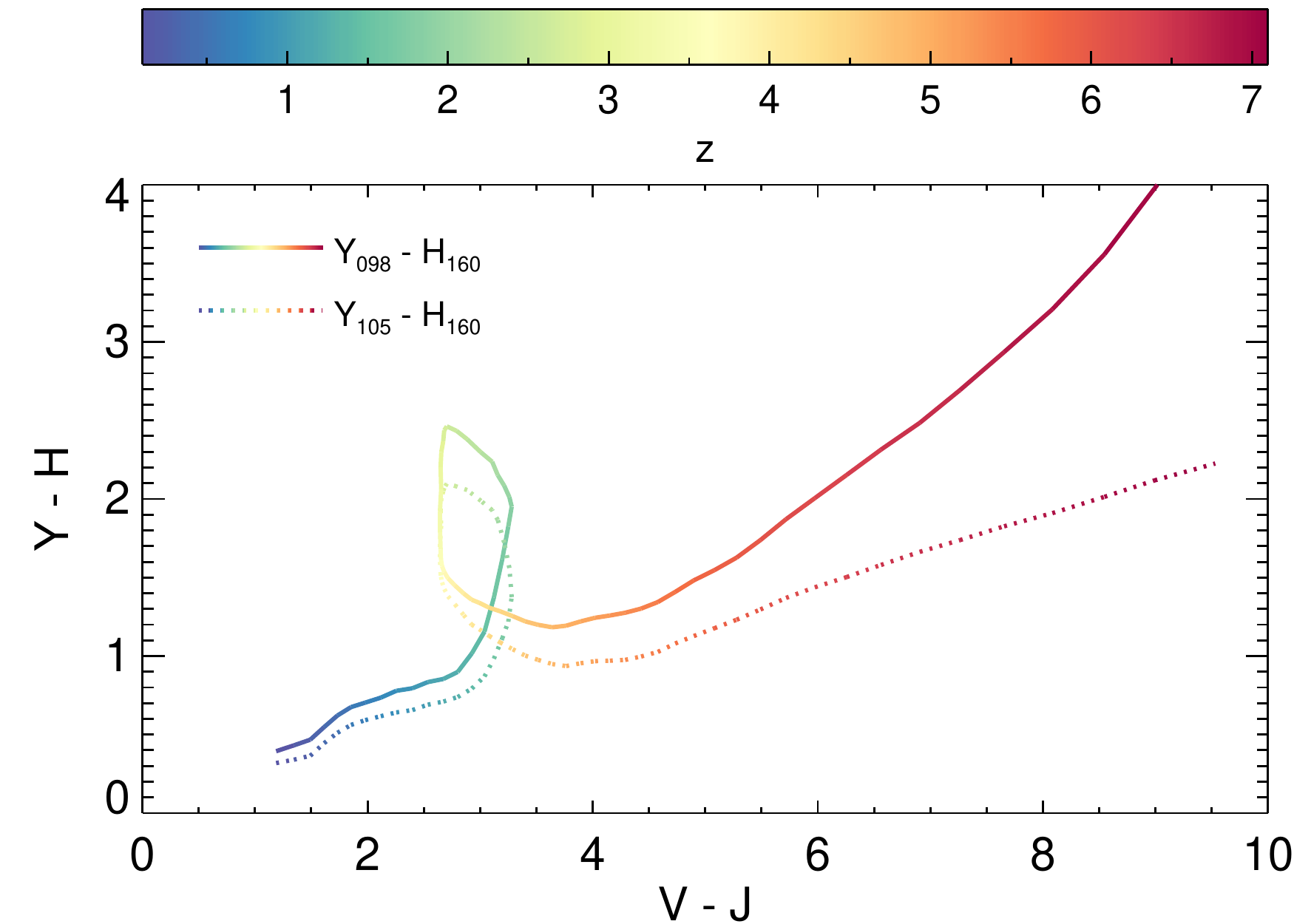}
	\caption{Plot of the $Y-H$ and $V-J$ colours changing with redshift for an old passive galaxy for filter sets with $Y_{098}$ and $Y_{105}$ filters. $Y_{098}-H_{160} > 1.5$ colour cut is an important preselection in the identification of $z\sim2$ candidates from the $YHVJ$ filter set used in the BoRG survey. $V-J$ colours are not directly used in our $z\sim2$ candidate selection.}
	\label{fig:colour_evolution}
\end{figure}

\begin{figure}
	\centering
	\includegraphics[width=\columnwidth]{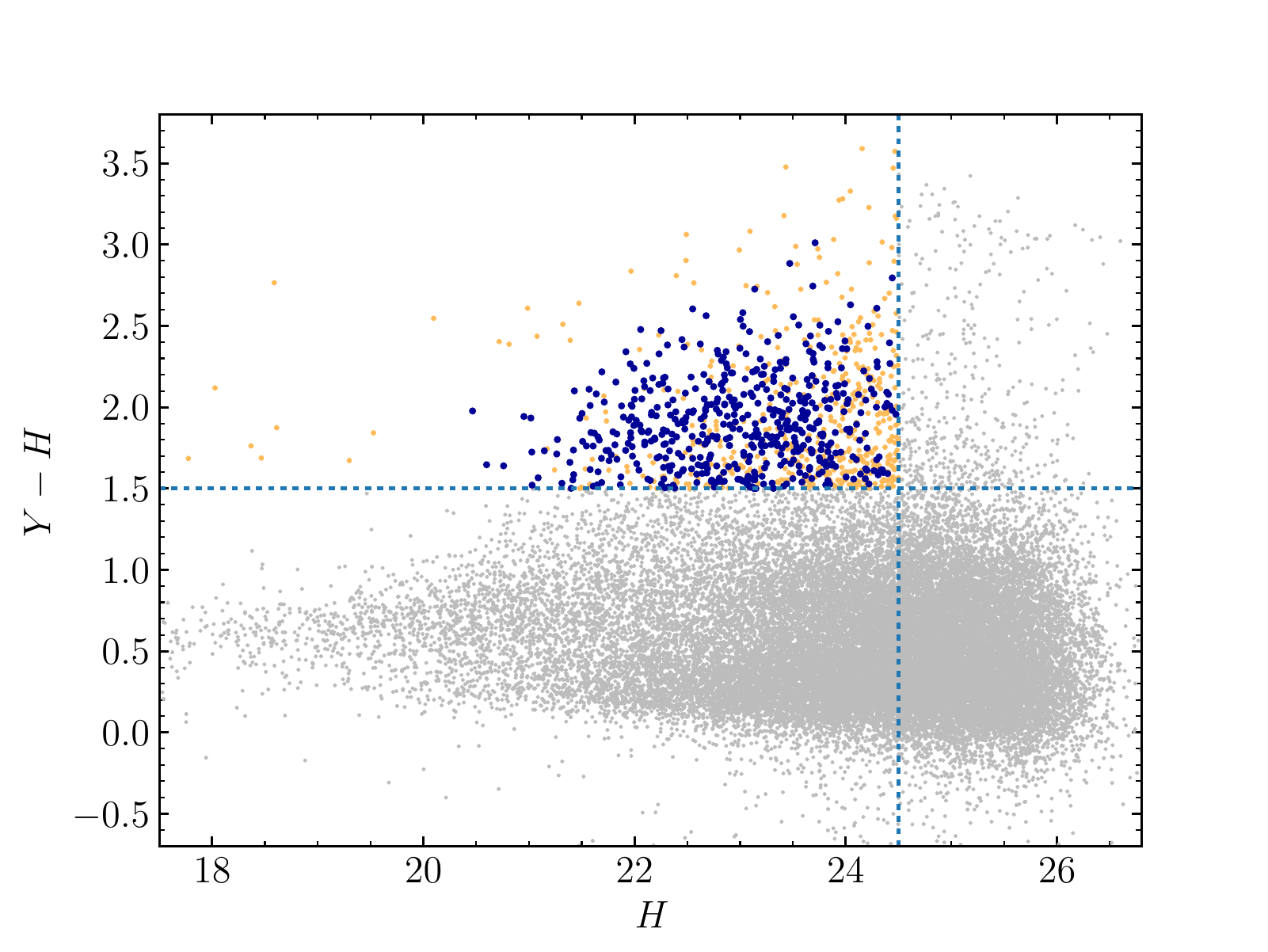}
	\caption{Colour--magnitude plot of all resolved sources in the BoRG[z8] sample. Dotted lines show magnitude cut-off ($H < 24.5$) and colour cut ($Y-H > 1.5$). Blue and yellow markers show sources passing these selection criterion. Blue markers depict visually confirmed candidates with photometric redshifts that passed all criteria and were thus included in our final sample ($1.5 < z < 2.5$, \texttt{chisq1} $< 1$ and \texttt{odds} $> 0.8$), while the yellow markers depict sources that failed one or more of these criteria and were excluded from our sample. Grey markers show all remaining sources detected with S/N > 8 that did not meet either or both of the magnitude cut-off or the $Y-H$ colour cut.}
	\label{fig:pass_fail_scatter}
\end{figure}

\subsection{Source Catalogs}
\label{source_detection}

The BoRG science data consist of reduced and aligned science images produced with \texttt{MultiDrizzle} \citep{Koekemoer03} as well as associated weight maps. From each weight map we create corresponding rms maps. In addition, a super-rms map is created by flagging all pixels that have zero weight in at least one of the filters. In this way we can include in the photometric analysis only pixels with complete spectral coverage, minimizing the impact of detector artifacts in the source catalogs and their associated photometric measurements. 

The rms maps are then normalised to account for correlated noise induced by \texttt{MultiDrizzle} \citep{Casertano00}. This is achieved by measuring the noise in randomly placed empty apertures (i.e., no overlap with detected sources) and comparing this to the noise value inferred from the rms map. Any disparity is corrected by applying a scaling factor to the rms map (see \citealt{Trenti11}).

Photometric zeropoints are set individually for each field by correcting the default \emph{HST} zeropoints for Galactic reddening along the line of sight, estimated from the $E(B-V)$ reddening factors in \citet{Schlafly11}.

Source catalogs in each filter were created for each field using \texttt{SExtractor} \citep{SExtractor} in dual-image mode with source detection performed in the $H_{160}$ band, with further pruning to include only sources with $S/N\ge 5$ in the detection image, where signal-to-noise ratios were calculated using isophotal flux and error (\texttt{FLUX\_ISO}/\texttt{FLUXERR\_ISO}). The \texttt{SExtractor} \texttt{CLASS\_STAR} parameter was used to exclude stars from the analysis, with sources having \texttt{CLASS\_STAR} > 0.8 being excluded. Given the non-uniform depth of the BoRG survey, we restricted the analysis to sources with \texttt{MAG\_AUTO} < 24.5, which ensures a magnitude-limited sample with high completeness (despite the fact that several fields are complete to fainter magnitudes).

\subsection{Redshift $z\sim2$ Candidate Selection}
\label{photometric_redshifts}

Near-infrared imaging with WFC3/IR using a $YHVz$ filter set has been shown to enable effective photometric selection of galaxies in the redshift range $1.5 < z < 3.5$. \citet{CameronE11} devised a $Y-H$ versus $V-z$ colour-colour selection to select for passive and star-forming galaxies in HUDF and GOODS-S, exploiting the presence of the rest-frame 4000 \AA{} break in this wavelength range. Their selection involves two criteria: star-forming galaxies are captured by selecting sources redder in $Y-H$ colour than in $V-z$ by a given amount, while the passive population (very red  in $V-z$) is captured with a simple cut, selecting all sources with $Y-H$ colour redder than a given value, determined by the $Y$-band filter used.

The limited filter set available in the BoRG survey compared to a legacy field survey prompted us to employ a one-colour selection followed by full application of photometric redshift to build our final sample. Figure \ref{fig:colour_evolution} shows the progression of $Y-H$ and $V-J$ colours of an example passive galaxy spectrum as it is moved to higher redshift. The knot in figure \ref{fig:colour_evolution} around redshift $z\sim2$, caused by the 4000 \AA{} break lying between the $Y$ and $H$ bands, is what we aim to capture with this $Y-H$ one-colour selection.

\subsubsection{BoRG[z8] candidate selection}
\label{sub:borgz8_candidates}

For the $Y_{098}$ filter used in the BoRG[z8] sample, we apply a cut of $Y_{098} - H_{160} > 1.5$, motivated by the work of \citet{CameronE11}, to select for passive galaxies as well as dust-reddened star-forming galaxies. Photometric redshift estimates were then obtained for sources passing this colour selection with \texttt{BPZ} \citep{Benitez00, Benitez04, Coe06}, a Bayesian Photometric Redshift software. Input catalogs were constructed for \texttt{BPZ} by compiling the isophotal magnitude (and the associated rms error) in each filter for each source (\texttt{MAG\_ISO} and \texttt{MAGERR\_ISO}).

Our final sample of $z\sim2$ candidates was selected based on parameters from the resultant \texttt{BPZ} output catalogs. Sources in the redshift range $1.5 < z < 2.5$ were selected based on the photometric redshift best-estimate (\texttt{zb}) output parameter. To ensure quality-of-fit for the redshift estimates, sources with $\texttt{chisq2} > 1$ were eliminated. Additionally, the \texttt{odds} measure (likelihood that \texttt{zb} is correct within 0.10) was used to exclude sources with multiple possible redshift solutions, with sources having $\texttt{odds} < 0.8$ being removed.

As a final vetting, all candidates were visually inspected and removed if they were judged to be photometric artifacts. From the initial 987 candidates across the full survey (BoRG[z8] and BoRG[z9]), 18 sources were removed due to non-equivocal identification as part of diffraction spikes created by bright stars. The colour-magnitude diagram in figure \ref{fig:pass_fail_scatter} shows the location of resolved sources passing the set of selection criteria in the BoRG[z8] sample.

\subsubsection{BoRG[z9] candidate selection: $Y_{105}-H_{160}$ Colour Cut}
\label{cut_compare}

The BoRG[z9] sample consists of a further 74 fields imaged on WFC3 with a slightly different filter set (see section \ref{data}). In particular, the BoRG[z9] sample employs a different $Y$-band filter, the wide band F105W filter, rather than the medium band F098M filter used in BoRG[z8]. Here we describe our approach to selecting a new $Y-H$ colour cut to mirror this change in filters. Other than this alternative colour cut, the same method was applied to the BoRG[z9] sample as for BoRG[z8], described in section \ref{sub:borgz8_candidates}. An overview of the results for the BoRG[z9] sample, considered in isolation to BoRG[z8], is presented in section \ref{borgz9_results}.

The efficiency of source selection and resultant sample redshift range depends strongly on the $Y-H$ cut used. Thus, the alternative $Y_{105}$ filter used for the BoRG[z9] sample requires a different magnitude colour cut to be applied. It is expected that a slightly lower cut-off would be required as the redder average wavelength of the F105W filter compared to the F098M filter would imply a particular $z\sim2$ galaxy would appear bluer in $Y_{105} - H_{160}$ colour than in $Y_{098} - H_{160}$.

\begin{figure}
	\includegraphics[width=\columnwidth]{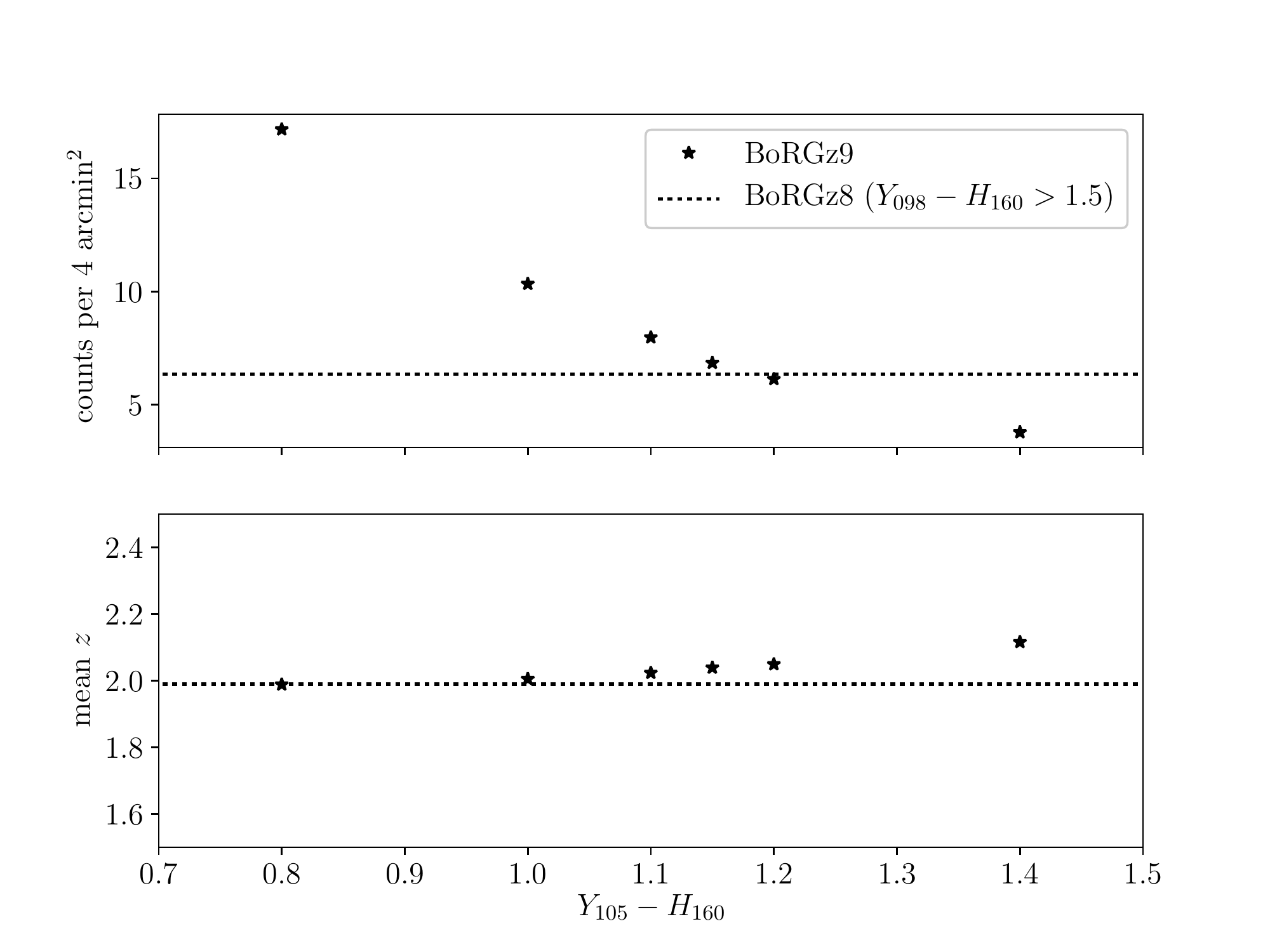}
	\caption{Mean scaled field counts and mean photometric redshift recovered from BoRG[z9] analysis with different $Y_{105} - H_{160}$ colour cuts. Horizontal dotted lines show benchmark values obtained from BoRG[z8] analysis using $Y_{098} - H_{160} > 1.5$. It is worth noting that these numbers include sources that would eventually be removed as artifacts by visual inspection. Given that only $\sim1\%$ of sources would be excluded on these grounds, it was considered that this  would not impact the outcome of this comparison.}
	\label{fig:colour_cuts}
\end{figure}

We considered two metrics for comparing the BoRG[z9] colour cut to that used in BoRG[z8]. Increasing the cut-off colour value would decrease mean number counts, but should increase the mean redshift derived for the population, due to the increased emphasis on redder objects. We reasoned that both of these quantities should be consistent between the two samples as they reflect properties of the galaxy populations being sampled. Figure \ref{fig:colour_cuts} summarises our investigation of how different $Y_{105} - H_{160}$ cuts impacted the mean counts per field and redshift distribution of the resultant sample.

Given the importance of Poisson statistics to the analysis, we eventually selected the cut that best reproduced the mean counts seen in BoRG[z8], adopting a cut-off of $Y_{105} - H_{160} > 1.2$. It is perhaps worth noting that although in varying this cut we influence the first moment of the distribution of counts (the mean), the derived galaxy bias depends primarily on the second moment (variance).

\subsubsection{Colour cut comparisons with borg\_0751+2917 and borg\_1209+4543}
\label{field_compare}

The fields borg\_0751+2917 and borg\_1209+4543 were serendipitously imaged in both BoRG[z8] and BoRG[z9], observed in parallel to separate revisits to the same primary target. Though in both cases the BoRG[z9] visits were discarded for our final analysis so as not to consider the same galaxy candidates twice in our set of field counts, they present a unique opportunity to compare the $Y-H$ colour cuts selected for each sample.

The $z\sim2$ candidate counts returned for the borg\_0751+2917 were eight and nine for the BoRG[z8] and BoRG[z9] samples respectively. The eight candidates counted in BoRG[z8] were all present in the BoRG[z9] final count along with one additional candidate, borgz9\_0751+2917\_317. In that case, the candidate in the BoRG[z8] pointing had been positioned near the edge of the field, affecting the detected shape of the object, ultimately resulting in the object not passing colour selection, with a colour of $Y_{098} - H_{160} = 1.44$.

The BoRG[z8] and BoRG[z9] samples both returned six $z\sim2$ candidates from the borg\_1209+4543 field. Four of these candidates were observed in both fields with each sample returning two candidates not observed in the other. Of these sources selected in one sample but not the other, one was near the $m_{AB} = 24.5$ threshold and was excluded in BoRG[z8] based on its magnitude, but selected in BoRG[z9]. The other three candidates were excluded from one sample based on the $Y-H$ colour cut, but selected in the other. In all three of these cases, the colour cut-off was well within the 1$\sigma$ uncertainty of the $Y-H$ value (given by \texttt{SExtractor}) for one of either the BoRG[z8] or the BoRG[z9] sample. In this sense, the fact that these sources were selected as $z\sim2$ candidates in one sample and not the other can be attributed to uncertainty in the photometry that is within the bounds of what is expected.

Between the two fields, a total of 14 sources passed colour selection in both samples. Among these sources twelve were selected as $z\sim2$ candidates by both samples and two were rejected by both samples (\textit{i.e.}, the two samples were in agreement in all cases). Of the twelve that were selected as $z\sim2$ candidates, the photometric redshifts obtained from each sample were in fairly good agreement. The largest photometric redshift discrepancy between the two samples was $|z_{\text{BoRG[z8]}}-z_{\text{BoRG[z9]}}| = 0.33$, with only four such discrepancies in excess of 0.20. The two rejected sources were excluded by both samples based on a low \texttt{odds} value, indicating the photometry was unable to distinguish between multiple possible redshift solutions in both cases.

These two fields unfortunately only make up a very small sample with which to compare how the selected colour cuts and photometry perform for the BoRG[z8] and BoRG[z9] samples. However, the limited crossover we have suggests that, within the bounds of the uncertainties of photometry for objects at this redshift, the two samples with the colour cuts as selected are yielding comparable galaxy populations.

\subsection{Normalization of Field Effective Area} 
\label{areascaling}

An accurate and unbiased comparison of number densities of $z \sim 2$ galaxies in different fields is critical to this analysis, however differences between the \citetalias{LopezSanjuan15} and \citetalias{Robertson10} methods mean that the normalization is implemented in different ways (these methods are outlined in more detail in section \ref{bias_calc}). Key to this normalization is an evaluation of the effective area sampled by each field. Several fields from the survey are combinations of two or more pointings and hence the angular sizes of these fields take a range of values. Moreover, large foreground objects may mask many $z \sim 2$ sources from detection. The presence of highly extended low-redshift galaxies and diffraction spikes from bright galactic stars must therefore also be taken into account.

For each field, an effective area was calculated from the \texttt{MultiDrizzle} weight maps and \texttt{SExtractor} segmentation check images to enable this comparison. From the weight map we summed the number of pixels that received a non-zero exposure time. These pixels were then excluded if they returned a non-zero value in the segmentation image, thus removing any pixels that were part of a detected object. Consequently, the presence of large objects obscuring a significant proportion of a field would be reflected in the derived area. 

In the \citetalias{LopezSanjuan15} method, the different fields are considered as volumes of varying size and the number counts used are the raw integer number counts obtained as per section \ref{sub:borgz8_candidates}. The co-moving volumes were calculated with the online calculator available from \citet{Trenti08} as described in section \ref{sig_DM}.

However, the \citetalias{Robertson10} method requires us to compare densities as though the fields sampled identical volumes. Thus for this method, the number counts for each field were scaled according to these areas such that they reflect a representative sampling of the galaxy density at that location in the sky. Additionally, key to the \citetalias{Robertson10} method for estimating galaxy bias is removal of the scatter due to Poisson variance from the distribution of galaxy counts. Due to the dependence of Poisson variance on the mean, it is important that scaling these galaxy counts preserves a mean that is representative of the survey sample volume. Thus, a scaling factor was applied to each field such that its count reflects an area of 4.41 arcmin$^2$, the median area calculated for the survey. This set of values was used in the calculation of the galaxy bias. The full field-by-field number counts can be found in Appendix \ref{field_counts}.

\subsection{Bias Calculation}
\label{bias_calc}

Galaxy bias is the ratio of the two-point correlation functions of galaxies and matter (equation \ref{eq:bias_def}). Assuming the bias in equation \ref{eq:bias_def} is scale-independent, the variance of galaxy ($\sigma^{2}_\textit{g}$, or ``cosmic variance'') and dark matter ($\sigma^{2}_\textit{DM}$) fields is related to the bias through 

\begin{equation}
\sigma^{2}_\textit{g} = b^{2}\sigma^{2}_\textit{DM}.
\label{eq:bias_cosvar_sigDM}
\end{equation}

We employ two independent methods to obtain $\sigma^{2}_\textit{g}$. The key quantity employed by both methods is the intrinsic scatter in a set of number counts from a large number ($N \gtrsim 50$) uncorrelated fields that represent discrete samplings from the underlying galaxy density field.

In sections \ref{source_detection} -- \ref{areascaling} we obtained a set of field-by-field $z\sim2$ candidate counts. The variance of this set, $\sigma^2_N$, is expected to have two components: one due to the intrinsic clustering on large scales ($\sim \sigma^2_{\text{g}}$), and one due to the Poisson statistics associated with taking discrete samplings from the galaxy density field.

The two methods differ subtly in their approaches toward disentangling these effects. The different methods are described in the following sections (\ref{sub:LS15} - \ref{sub:R10})

\subsubsection{Robertson Method}
\label{sub:R10}

The \citet{Robertson10} method is based in the assumption that in the absence of clustering, our galaxy counts would fit a Poisson distribution with $\sigma^2_{\text{Poisson}} = \bar{N}$. In fact, our counts exhibit an excess scatter, with $\sigma^2_N > \bar{N}$, attributed to large scale clustering (see Figure \ref{fig:initial_field_counts}). Following \citetalias{Robertson10}, we calculate this excess to yield $\sigma^2_{\text{g}}$,

\begin{equation}
\sigma^2_{\text{g}} \approx \frac{\sigma^{2}_{N} - \bar{N}}{\bar{N}^{2}}.
\label{eq:sig_g}
\end{equation}

Given we expect $\sigma^{2}_\textit{g}$ and $\sigma^{2}_\textit{DM}$ to be related linearly via the bias, we can use this value of $\sigma^{2}_\textit{g}$ and a value of $\sigma^{2}_\textit{DM}$ predicted from theory (see section \ref{sig_DM}) to estimate the galaxy bias.

\begin{equation}
b^{2} =  \frac{\sigma^{2}_\textit{g}}{\sigma^{2}_\textit{DM}} \approx \frac{\sigma^{2}_{N} - \bar{N}}{\bar{N}^{2}\sigma^{2}_{DM}}.
\label{eq:bias}
\end{equation}

\subsubsection{L\'opez-Sanjuan Method}
\label{sub:LS15}

The \citet{LopezSanjuan15} method differs in that it obtains a measure of the intrinsic scatter by fitting a log-normal distribution to the number densities after estimating the Poisson shot noise for each volume. 

The variance in galaxy counts due to cosmic variance is defined as 

\begin{equation}
\sigma^2_{\text{g}} = \dfrac{\left\langle N^2 \right\rangle - \left\langle N \right\rangle^2}{\left\langle N \right\rangle^2} - \dfrac{1}{\left\langle N \right\rangle}.
\label{eq:cosvar}
\end{equation}

The $1/\left\langle N \right\rangle$ term is to correct for Poisson noise. In the \citetalias{LopezSanjuan15} method, Poisson shot noise is estimated to be 

\begin{equation}
\sigma_{P,j} = \sqrt{N_j}/V_j
\label{eq:shot_noise}
\end{equation}

where $N_j$ and $V_j$ are the number counts and volumes of sub-field $j$. Using the maximum likelihood estimator (MLE) presented in \citet{LopezSanjuan14}, a log-normal distribution is fit to our field densities (number counts obtained in section \ref{sub:borgz8_candidates} scaled according to co-moving volumes derived in line with section \ref{sig_DM}) and the median number density $\bar{n}$ and intrinsic dispersion $\sigma_{\text{int}}$ for the sample are obtained. 

Applying equation \ref{eq:cosvar} to $P_{\text{LN}}(n)$, \citetalias{LopezSanjuan15} conclude that the observed relative cosmic variance is

\begin{equation}
\sigma_{g}^2 = e^{\sigma^2_{\text{int}}} - 1.
\label{eq:observed_cosvar}
\end{equation}

We then obtained a value for galaxy bias $b_g$ by applying equation \ref{eq:bias_cosvar_sigDM} using a value for $\sigma^2_{DM}$ obtained as described in section \ref{sig_DM}.

\subsection{Value for variance of the matter field ($\sigma^{2}_{DM}$)}
\label{sig_DM}

In calculating the galaxy bias from the methods described, we consider $\sigma^{2}_{DM}$ over the volume spanned by the median field area (2.1 arcmin $\times$ 2.1 arcmin) across the redshift interval $\bar{z} - \sigma_{z} < z < \bar{z} + \sigma_{z}$, where $\bar{z}$ and $\sigma_{z}$ are the mean and standard deviation of the photometric redshifts obtained across the whole sample. This corresponds to comoving dimensions of approximately $2.3 \times 2.3 \times 560$ Mpc$^{3} h^{-3}$ for the BoRG[z8] sample (assuming a field of view of 4.41 arcmin$^2$). 

We calculate $\sigma^{2}_{DM}$ using the online calculator available from \citet{Trenti08}, assuming $\sigma_8 = 0.815$ \citep{Planck16}. We recalculate a new value for each survey subsample considered, using the precise redshift distribution obtained for that subsample. A summary of the volumes considered and values obtained for each of the subsamples in our analysis can be found in Table~\ref{tab:sigma_dm}.

\begin{table}
	\centering
	\caption{Values used for variance of matter field in bias calculation. All values are calculated with \citet{Trenti08} using the available online calculator. Comoving dimensions of the pencil beam geometry adopted are given. A value of $\sigma_8 = 0.815$ is assumed \citep{Planck16}.}
	\label{tab:sigma_dm}
	\begin{tabular}{lccc}
		\hline
		Sample & $z$ interval & $\sigma^{2}_\textit{DM}$ & Comoving Dimensions \\
		& & & Mpc $h^{-1}$ \\
		\hline
		Combined & 2.02 $\pm$ 0.26 & 0.0118 & $2.31 \times 2.31 \times 566$\\
		\hline
		BoRG[z8] & 1.99 $\pm$ 0.25 & 0.0125 & $2.29 \times 2.29 \times 549$\\
		BoRG[z9] & 2.05 $\pm$ 0.27 & 0.0112 & $2.23 \times 2.23 \times 574$\\
		\hline
		Early type & 1.89 $\pm$ 0.27 & 0.0120 & $2.22 \times 2.22 \times 615$\\
		Late type & 2.11 $\pm$ 0.22 & 0.0139 & $2.37 \times 2.37 \times 452$\\
		\hline
	\end{tabular}
\end{table}

\subsection{Bootstrapping Error Analysis}
\label{bootstrapping}

The \citet{LopezSanjuan14} MLE used for the \citetalias{LopezSanjuan15} method provides 1$\sigma$ error estimates. For values calculated from the \citetalias{Robertson10} method, we quantify the uncertainty with a Monte-Carlo bootstrapping resampling method. In this method, counts from the original data set are selected at random with repetition allowed to construct a synthetic data set of length $N_{\textit{fields}}$ with which a new bias calculation could be performed (section \ref{bias_calc}). This bias calculation is repeated 1000 times for newly generated synthetic sets, yielding a distribution of bias values. The 1$\sigma$ uncertainty of this distribution was adopted as the 1$\sigma$ uncertainty for the calculated bias value.

\section{Results and analysis}
\label{results}

In this section we describe the candidate sample obtained as well as the bias values obtained from each of the \citetalias{LopezSanjuan15} and \citetalias{Robertson10} methods. Since candidates in the BoRG[z9] sample were selected using a different colour cut (see section \ref{cut_compare}), we initially consider each of the BoRG[z8] and BoRG[z9] samples in isolation (sections \ref{borgz8_results} and \ref{borgz9_results}). After establishing the bias results are broadly comparable in both samples (as expected), we combine the two samples and compute overall bias values with the both methods (section \ref{final_value}). We then divide the candidates into early-type and late-type sub-samples to examine the dependence of the bias on spectral type (\ref{spectral_type_bias}). It is worth noting that the two methods give different values, with disagreements in excess of 1$\sigma$ in some cases. We discuss the differences between the two methods in section \ref{sub:method_comparison}.

\subsection{BoRG[z8] Sample}
\label{borgz8_results}

We identify 490 $z\sim2$ candidates across the 69 fields in the BoRG[z8] survey with a mean count of 7.1 candidates per field and a field-to-field variance of 13.1. 

All but the most luminous candidate have $H$-band magnitudes of $20.5 < m_{H} < 24.5$, with the brightest candidate having $m_{H} = 20.47$. Number counts increase steadily with decreasing luminosity to a magnitude of $m_H = 23.0$ with $23.0 < m_H < 23.5$ being the most populated magnitude bin. A higher proportion of sources failing photometric redshift selection at magnitudes fainter than $m_H = 23.5$ leads to a decrease in number counts out to $m_H = 24.5$, despite a proportionally high number of sources passing $Y-H$ colour selection across this magnitude range.

\begin{figure}
	\includegraphics[width=\columnwidth]{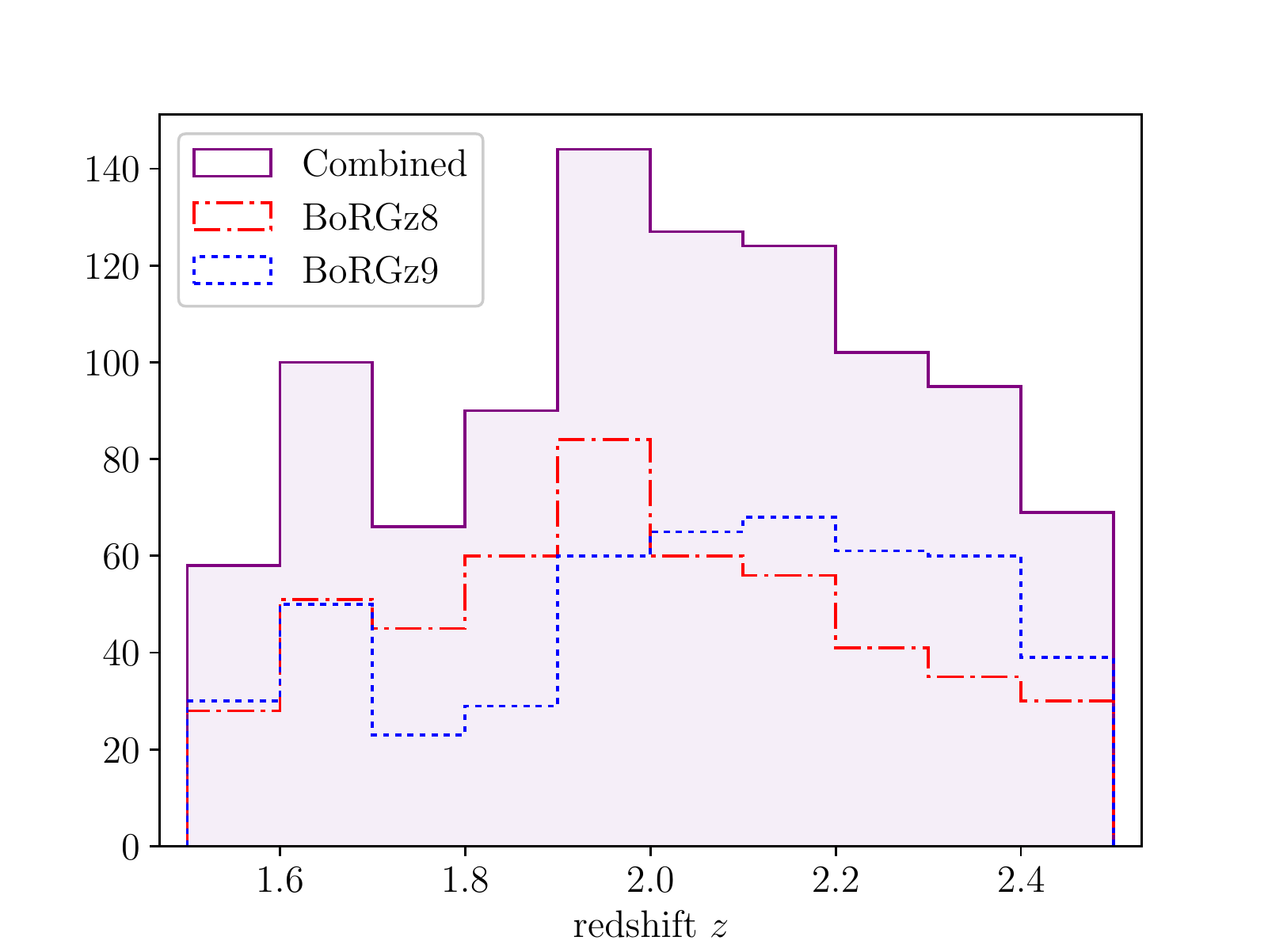}
	\caption{Distribution of best photometric redshifts as determined with BPZ of $z \sim 2$ candidates passing colour cut and BPZ quality-of-fit criteria. Dash-dot line shows the initial BoRG[z8] sample with the $Y_{098}-H_{160} > 1.5$ colour cut, while dotted line shows the BoRG[z9] sample using a $Y_{105}-H_{160} > 1.2$ colour cut. The solid line depicts the final redshift distribution after combining the two sets.}
	\label{fig:zb_dist}
\end{figure}

Approximately $60\%$ of candidates passing all selection criteria had $S/N$ in the range $5 < S/N < 50$, while only $9\%$ had $S/N > 100$. Sources for which we obtained photometric redshifts were selected to fit the range $1.5 < \texttt{zb} < 2.5$ with a resultant mean redshift of $\bar{z}=1.99$ and standard deviation of $\sigma_z=0.25$ across the sample, as seen in figure \ref{fig:zb_dist}. Quality-of-fit selection was conducted based on \texttt{odds} and \texttt{chisq2} output parameters from \texttt{BPZ}. The criteria $\texttt{odds} > 0.8$ and $\texttt{chisq2} < 1.0$ were selected for with $86\%$ of candidates having $\texttt{odds} > 0.9$ and $82\%$ with $\texttt{chisq2} < 0.50$.

\begin{figure}
	\centering
	\includegraphics[width=\columnwidth]{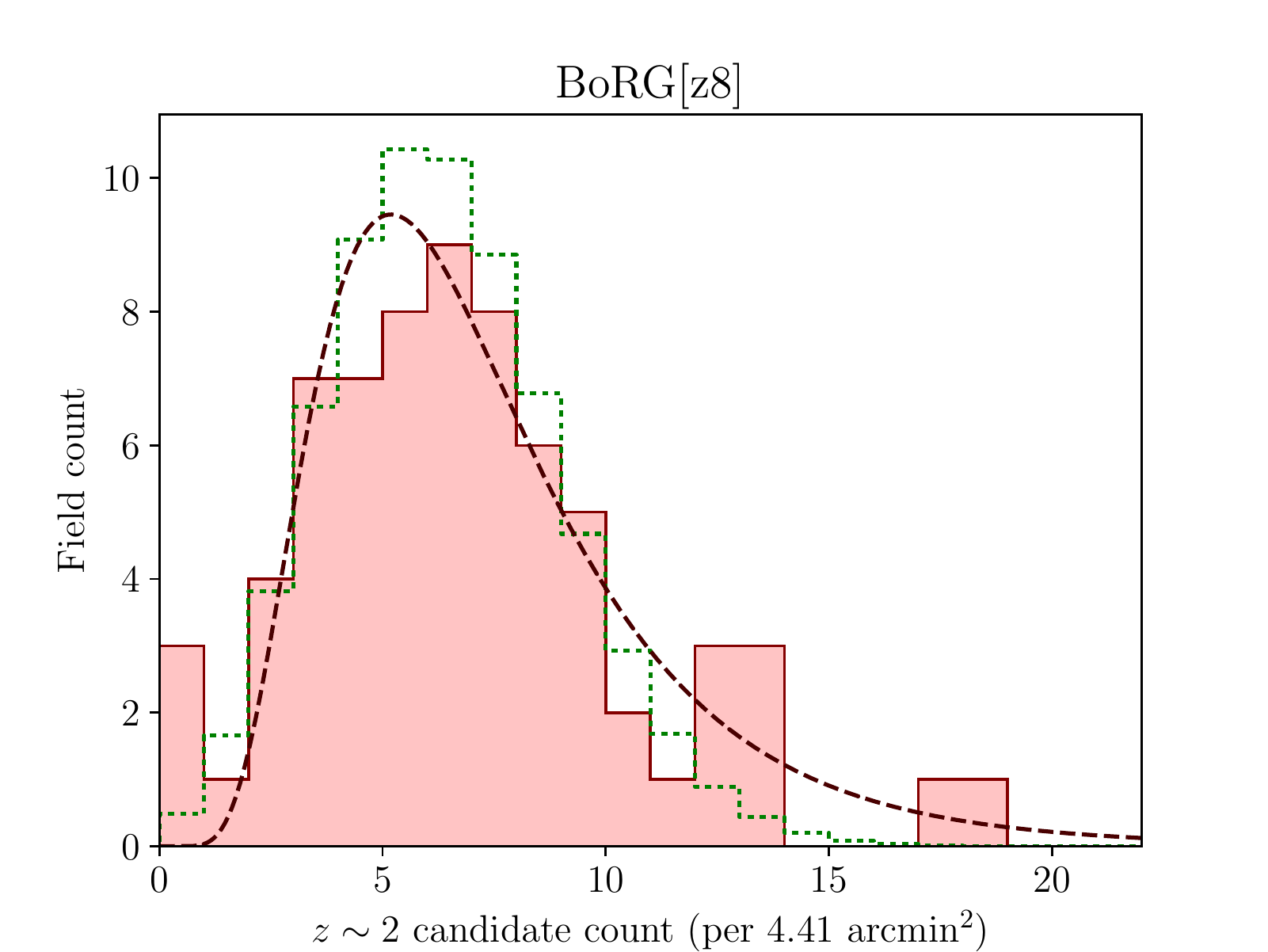}
	\includegraphics[width=\columnwidth]{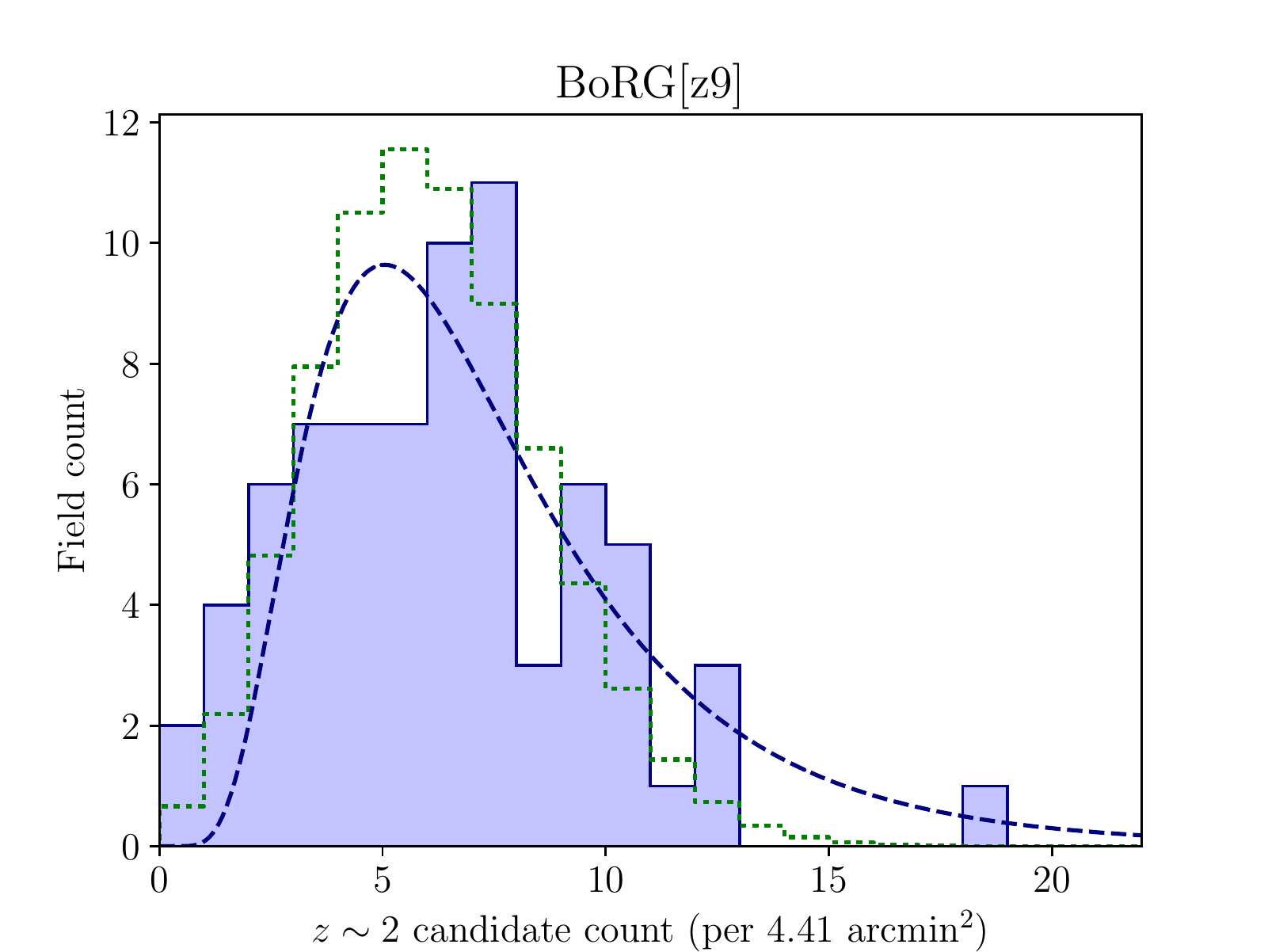}
	\caption{Distributions of $z\sim2$ candidate counts across fields in the BoRG[z8] and BoRG[z9] subsamples of the BoRG survey. The scatter in these distributions is expected to have two components: one due to Poisson uncertainty and one due to cosmic variance. The green dotted histogram shows a Poisson distribution with the same mean and normalisation as the set of field counts. In both the BoRG[z8] and BoRG[z9] samples, a broadening of set of field counts beyond Poisson uncertainty is observed, as expected. The \citetalias{Robertson10} method derives an estimate of the galaxy bias from this simple excess variance. The \citetalias{LopezSanjuan15} method fits a log-normal distribution to the densities (dashed line shows log-normal fit to scaled field counts for illustration). Note that the plotted line is not the distribution used for calculating the bias, that involves fitting a normal to log densities. The intrinsic scatter, or ``cosmic variance'' is derived from the log-normal fit, affording a value for the galaxy bias. These methods are described in more detail in section \ref{bias_calc}.}
	\label{fig:initial_field_counts}
\end{figure}

Applying the methods described in sections \ref{bias_calc} -- \ref{bootstrapping} to the set of counts obtained from the BoRG[z8] sample yields a preliminary value for the galaxy bias of $b \approx 2.88 \pm 0.60$ derived from the \citetalias{LopezSanjuan15} method or $b \approx 3.20 \pm 0.68$ as derived from the \citetalias{Robertson10} method (refer to section \ref{sub:method_comparison} for a discussion of the difference in values obtained from the two methods). Figure \ref{fig:initial_field_counts} shows histograms of the number of fields with given numbers of candidates.

\subsection{BoRG[z9] Sample}
\label{borgz9_results}

The BoRG[z9] sample consists of 74 fields, including the 2 fields identified as BoRG[z8] revisits (see section \ref{field_compare}), totalling 72 new unique fields. We identify 494 $z\sim2$ candidates across these 74 fields at a slightly lower mean of 6.7 candidates per field, but with a higher field-to-field variance of 14.9.

The BoRG[z9] sample shows no significant deviations from BoRG[z8] in terms of the magnitude and signal-to-noise ratio distributions it returns, with the brightest candidate having a magnitude of $m_H = 20.81$.

Photometric redshifts from \texttt{BPZ} returned a mean redshift of $\bar{z}=2.05$ and a standard deviation of $\sigma_z=0.27$, both values slightly higher than those from BoRG[z8]. The distribution of quality-of-fit (\texttt{odds} and \texttt{chisq2}) values showed no significant change from that seen in BoRG[z8].

We calculated a value for the galaxy bias for BoRG[z9] considered in isolation as a final comparison of the candidate populations of the two samples. Following the procedure outlined in section \ref{bias_calc} -- \ref{bootstrapping}, we obtained a new $\sigma^{2}_{DM}$ value, to reflect the subtly different redshift distribution observed, before obtaining a galaxy bias of $b \approx 2.93 \pm 0.61$ from the \citetalias{LopezSanjuan15} method and $b \approx 3.94 \pm 1.02$ from the \citetalias{Robertson10} method. A slightly higher field-to-field variance in the scaled counts in the BoRG[z9] sample results in a higher estimate for the bias when compared to the BoRG[z8] results with both methods. However, both methods yield values that are within $1\sigma$ uncertainty of the BoRG[z8] value. Encouragingly, the uncertainties for the two values derived from the \citetalias{LopezSanjuan15} method are very similar, as expected due to a similar number of fields being present in each sample. These similar values further suggest to us that the two samples are returning comparable candidate populations.

\subsection{Combined Sample}
\label{combined_sample_results}

\subsubsection{Overview}

In order to implement the two methods on as large a data set as possible, we desired a combined sample made up of number counts from fields across both the BoRG[z8] and BoRG[z9] samples. The selection process for $z\sim2$ candidates is identical in both samples, save for different cutoff values in the $Y-H$ colour cut as discussed in sections \ref{cut_compare} -- \ref{field_compare}. Satisfied that these two colour cut values were returning comparable candidate populations, we combined the sets of field counts obtained for each sample into one combined set with which we conduct our final bias analysis.

This yielded a final data set covering 141 fields (we ignore the BoRG[z9] visits to the two doubly imaged fields discussed in section \ref{field_compare}). From these fields our sample consists of a total of 969 candidates, with a mean of 6.9 candidates per field and a variance of 14.3 across the complete sample.

\begin{figure}
	\centering
	\includegraphics[width=\columnwidth]{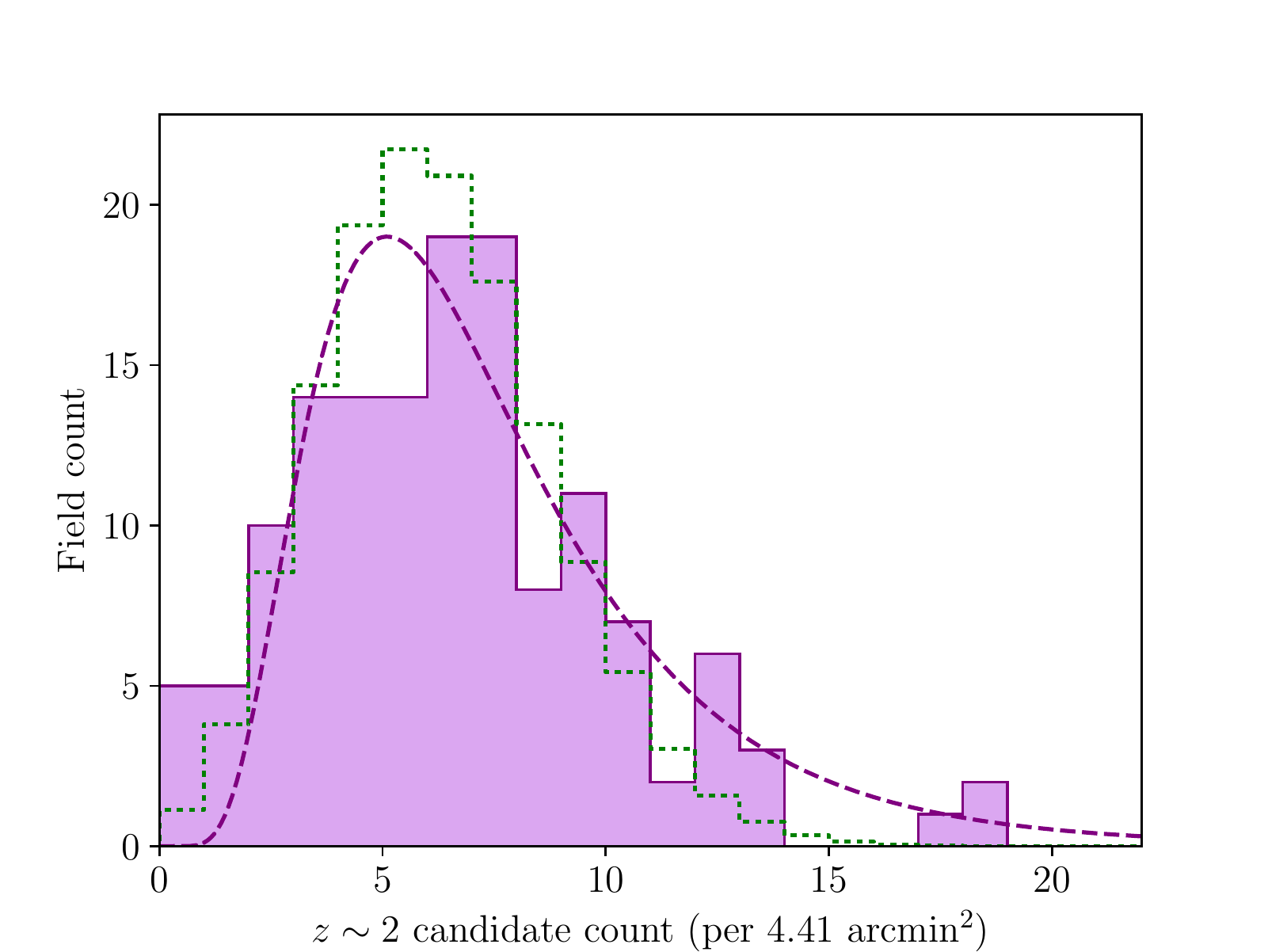}
	\caption{Distribution of number of counts per field (rescaled to area of 4.41 arcmin$^{2}$) after combining counts from the BoRG[z8] and BoRG[z9] samples. Dotted histogram shows Poisson distribution with same mean and normalisation. Excess scatter due to cosmic variance is clearly seen in the distribution of counts. Dashed line shows the best-fit log-normal to the scaled field counts shown.}
	\label{fig:combo_field_counts}
\end{figure}

The distribution of photometric redshifts obtained across the complete sample had a mean of $\bar{z}=2.02$ and a standard deviation of $\sigma_z=0.26$. The distribution of magnitude, signal-to-noise ratio, and photometric redshift quality-of-fit diagnostics among the candidates show no significant deviation from those described for the BoRG[z8] sample in section \ref{borgz8_results}. The distribution of field counts obtained for the combined sample is shown in figure \ref{fig:combo_field_counts}.

\subsubsection{Overall bias value}
\label{final_value}

\begin{figure}
	\includegraphics[width=\columnwidth]{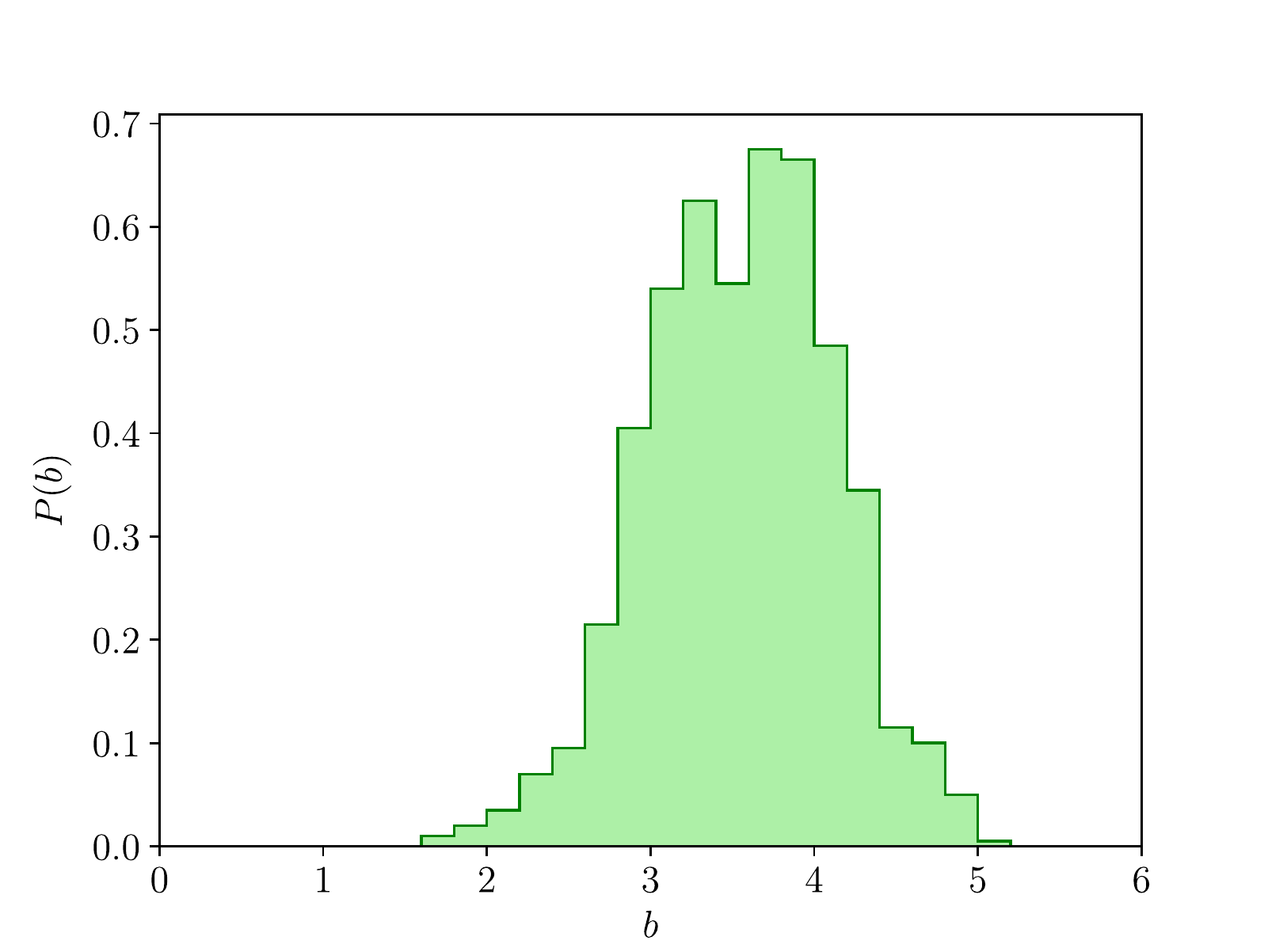}
	\caption{PDF of galaxy bias values obtained from Monte-Carlo Bootstrapping error analysis for the final value from the combined analysis using the \citetalias{Robertson10} method.}
	\label{fig:bias_error}
\end{figure}

Applying the \citetalias{LopezSanjuan15} method (see sections \ref{bias_calc} -- \ref{bootstrapping}) to the combined set of field counts yielded a value of $b \approx 2.94 \pm 0.41$ for our complete sample, while the \citetalias{Robertson10} method arrived at a value of $b \approx 3.63 \pm 0.57$ (probability distribution of the \citetalias{Robertson10} bias value obtained with Monte-Carlo bootstrapping is shown in figure \ref{fig:bias_error}). No previous studies have measured the bias of galaxies at $z\sim2$ using the same selection criteria as applied in this analysis, making direct comparisons to the literature somewhat difficult. \citet{Wake11} explored the evolution of bias of galaxies of different stellar masses between $z\sim1-2$ using photometric redshift estimations applied to the NEWFIRM Medium Band Survey. The bias for the lowest stellar mass bin of the $\bar{z} = 1.9$ sample identified in this survey was measured at $3.30^{+0.45}_{-0.23}$, broadly in line with our measurements. However, this corresponds to the NMBS magnitude limit of $K = 22.8$ which is slightly brighter than the cut-off for our sample.

\subsubsection{Dependence on spectral type}
\label{spectral_type_bias}

\begin{figure}
	\centering
	\includegraphics[width=\columnwidth]{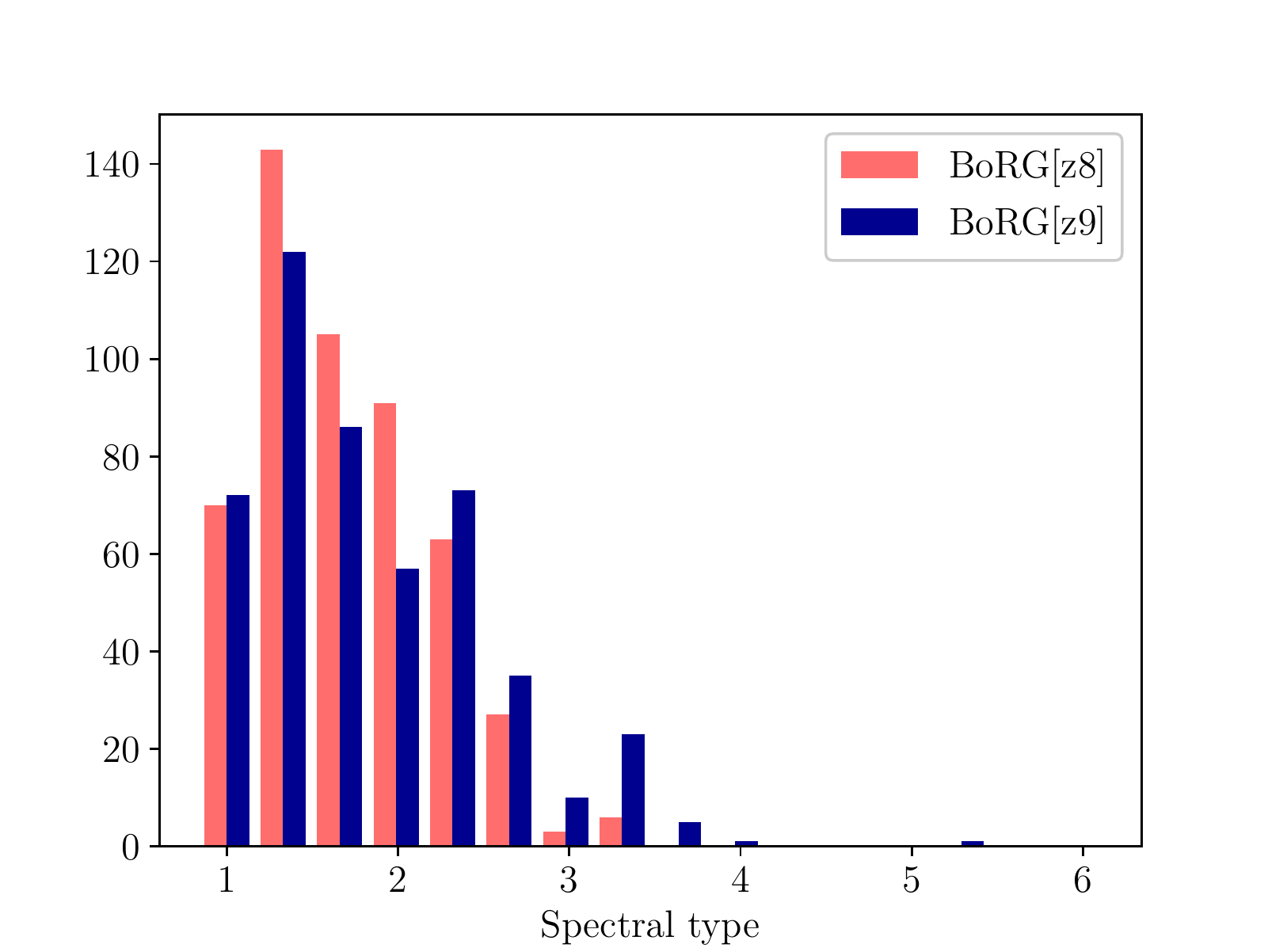}
	\caption{Distribution of spectral templates used by \texttt{BPZ} for $z\sim2$ candidates in each of the BoRG[z8] and BoRG[z9] samples. Early type galaxies are represented by a template of 1. All templates greater than 1 correspond to late-type and star-forming galaxies with the higher type numbers representative of bluer spectra. \texttt{BPZ} discretely interpolates templates, thus fractional type values correspond to sources assigned two templates at a one-to-two weighting. We take candidates with $\texttt{tb} \leq 1.33$ as early-type and $\texttt{tb} \geq 1.67$ as late-type. For a full description of these templates, the reader is referred to \citet{Coe06}.}
	\label{fig:type_hist}
\end{figure}

As a further exploration of the applicability of these methods, we divided our sample into early and late type galaxies based on the \texttt{tb} spectral type parameter which describes the SED template fit used by \texttt{BPZ} in determining the photometric redshift. Figure \ref{fig:type_hist} shows the distribution of spectral type parameters among candidates selected in the BoRG[z8] and BoRG[z9] samples. Over the combined sample this amounted to 390 early type and 585 late type candidates with mean photometric redshifts of $\bar{z}_\textit{early} = 1.89$ and $\bar{z}_\textit{late} = 2.11$ respectively. Considering these samples separately yields galaxy biases of $b_\textit{{early}} \approx 4.06 \pm 0.67$ and $b_\textit{late} \approx 2.98 \pm 0.98$ when calculated with the \citetalias{Robertson10} method.

Many previous studies of the clustering of passive and star-forming galaxies at redshift $z\sim2$ have utilised the well-established $BzK$ photometric selection method \citep{Daddi04}. Although it is worth noting that this selection yields a slightly different population to that yielded by our selections, clustering studies of passive and star-forming galaxies identified using the $BzK$ method with MUSYC \citep{Blanc08} measured a bias of $b_{\text{early}} = 3.27^{+0.46}_{-0.47}$ for a sample of passive galaxies at $\bar{z} = 1.58$ and a bias of $b_{\text{late}} = 2.93^{+0.59}_{-0.60}$ for a sample of star-forming candidates at $\bar{z} = 1.78$. Notwithstanding the differences in the construction of these galaxy samples, these values are broadly in line with our measurements, demonstrating that the \citetalias{Robertson10} method yields reasonable bias estimates with sufficient pointings. 

\begin{table}
	\centering
	\caption{Final values for the galaxy bias combining data from both BoRG[z8] and BoRG[z9] samples with the individual values from the BoRG[z8] sample for reference. The second column shows values derived with the method outlined in \citetalias{LopezSanjuan15}, fitting a log-normal distribution to the number counts. The third column contains values derived from the \citetalias{Robertson10} method in which the distribution is assumed to be Poisson plus scatter with the cosmic variance being derived from this excess scatter. Mean photometric redshifts for each sub-sample are given in the final column.}
	\label{tab:bias_results}
	\begin{tabular}{lccc}
		\hline
		Data set & $b_g$ (LS15) & $b_g$ (R10) & Mean $z$ \\
		\hline
		Overall & 2.94 $\pm$ 0.41 & 3.63 $\pm$ 0.57 & 2.02\\
		\hline
		BoRG[z8] & 2.88 $\pm$ 0.60 & 3.20 $\pm$ 0.68 & 1.99\\
		BoRG[z9] & 2.93 $\pm$ 0.61 & 3.94 $\pm$ 1.02 & 2.05\\
		\hline
	\end{tabular}
\end{table}

\begin{table}
	\centering
	\caption{Bias values for early and late spectral type sub-samples of the combined BoRG sample.}
	\label{tab:bias_results_type}
	\begin{tabular}{lcr}
		\hline
		Spectral type & $b$ & Mean $z$ \\
		\hline
		Early & 4.06 $\pm$ 0.67 & 1.89\\
		Late & 2.98 $\pm$ 0.98 & 2.11\\
		\hline		
	\end{tabular}
\end{table}

\subsection{Comparison of the Two Methods}
\label{sub:method_comparison}

While the two methods arrived at values that were broadly in line with the values we expect from literature measurements, they did not agree within 1$\sigma$ with each other for the combined sample. We discuss the reasons behind this disagreement here.

The key difference between the BoRG survey used here and ALHAMBRA survey originally used by \citetalias{LopezSanjuan15} is the volume sampled by each survey. The ALHAMBRA consisted of 8 fields with areas ranging from $720 - 1500$ arcmin$^2$ with the survey covering a total of 2.381 deg$^2$ (8571.6 armin$^2$). For the bias analysis, their sample was further divided, with each field being divided into sub-fields around $\sim$180 arcmin$^2$ in area (resulting in a total of 48 sub-fields). This contrasts starkly with the BoRG survey in which the total area sampled is approximately $\sim$700 arcmin$^2$ with the median field area being 4.41 arcmin$^2$.

\begin{figure}
	\centering
	\includegraphics[width=\columnwidth]{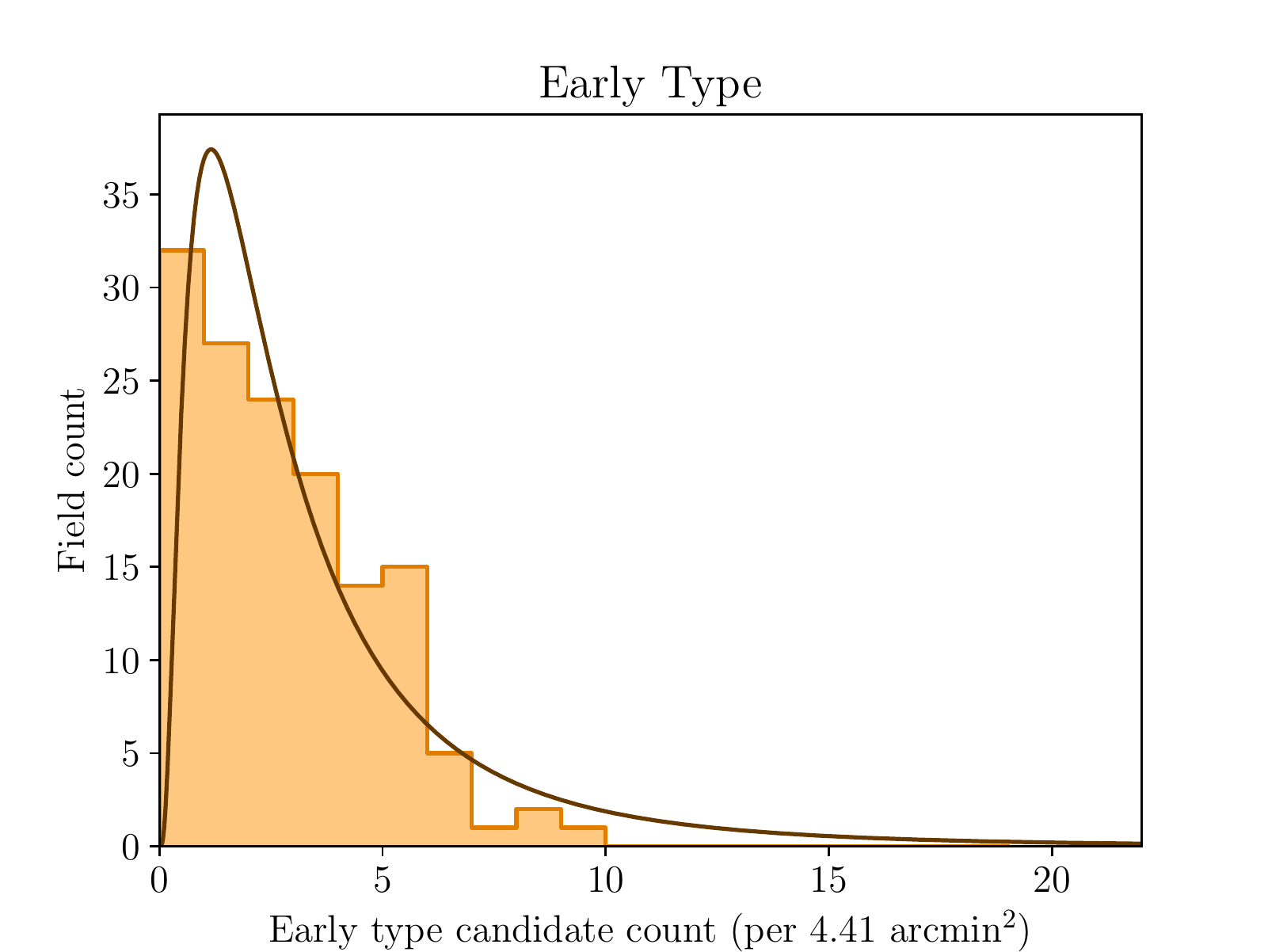}
	\includegraphics[width=\columnwidth]{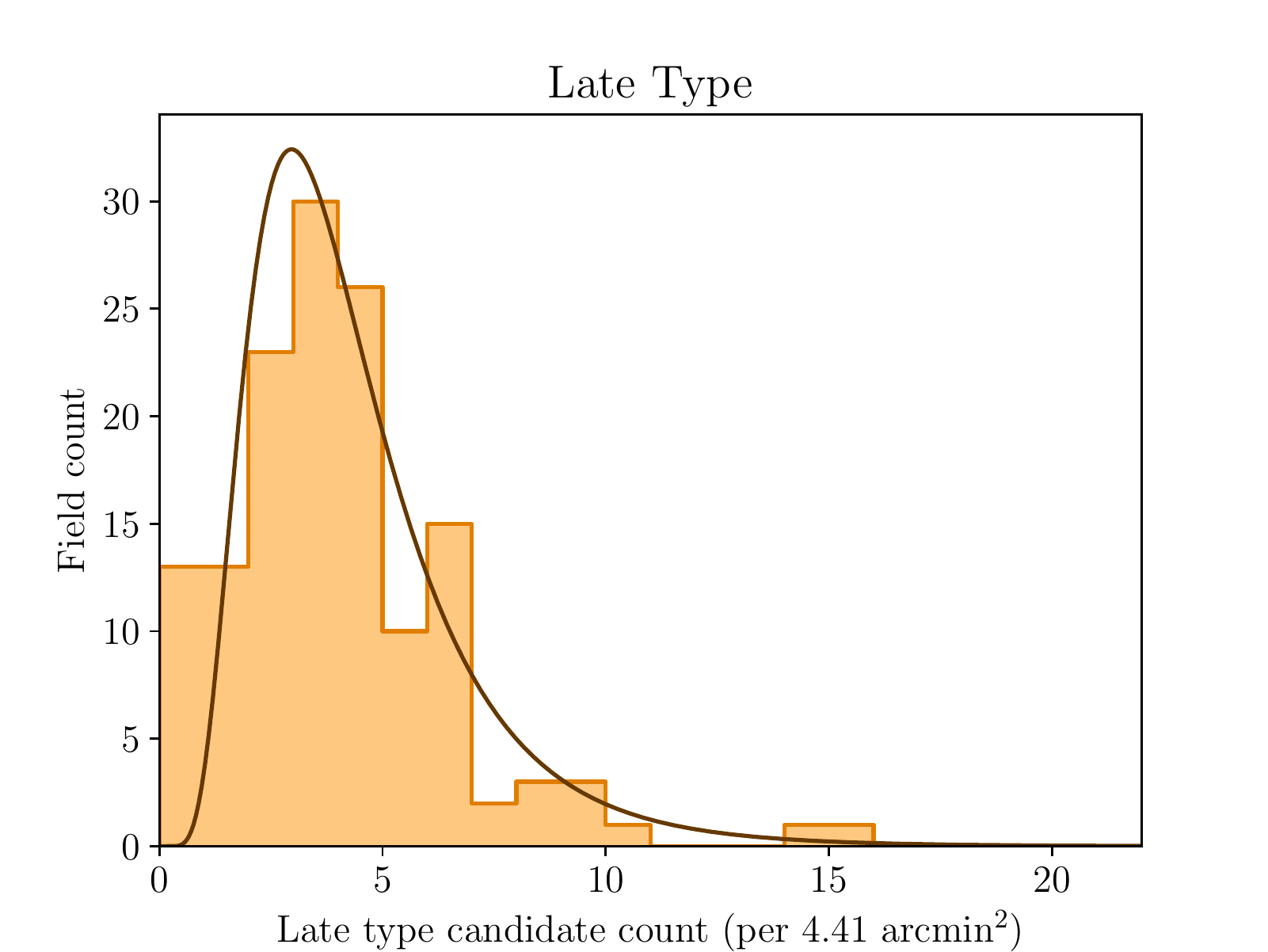}
	\caption{Histograms showing the distribution of candidate counts per field (scaled to 4.41 arcmin$^2$) after dividing sample into early- and late-type candidates (based on \texttt{tb} parameter from \texttt{BPZ} output catalog). The solid curves show the best-fit log-normal distributions to these counts. In the top panel, $0 \leq N_{\text{early}}<1$ is the most populated bin in the early-type distribution, and thus the log-normal fit presents significant systematic discrepancies that lead to an underestimation of the distribution variance. (Note: analysis is not conducted with the fits shown, these are for illustrative purposes only. Analysis is conducted by fitting a normal to log-densities with the \citet{LopezSanjuan14} MLE.)}
	\label{fig:type_counts}
\end{figure}

An important consequence of this is that the number counts obtained in our sample are much lower and in particular, the scatter within our sample suffers from a much higher contribution from Poisson shot noise. While we found that the \citetalias{LopezSanjuan15} method still yielded reasonable results for our complete combined sample, when we divided our sample into two sub-samples based on spectral type (section \ref{spectral_type_bias}), the method appeared to break down ($b_{\text{early}} = 1.39 \pm 0.95$ and $b_{\text{late}} = 2.57 \pm 0.52$ with the \citetalias{LopezSanjuan15} method). This is presumably due to the fact that 390 early type candidates across 141 BoRG fields (at a mean number count of 2.77 candidates per field) does not provide a sample with sufficiently high number counts in each field to adequately describe the distribution with a log-normal fit. Indeed, the most populated bin for early type candidates is $0 \leq N_{\text{early}} < 1$ (using integer bin edges --- see figure \ref{fig:type_counts}). This conclusion is supported by the fact that the \citetalias{LopezSanjuan15} method yielded a much more reasonable bias value for the late-type population where number counts were higher (585 late-type candidates across the sample; only three fields registering no late-type candidates).

The \citetalias{Robertson10} method does not enforce a distribution on the set of number counts, and thus it has the advantage of being more robust in samples with low number counts where Poisson shot noise has a more significant impact. Considering we obtained reasonable results with the both methods for the full sample, but were only able to obtain reasonable results with the \citetalias{Robertson10} method after dividing the sample based on spectral type, we suggest that our sample lies somewhat near the boundary of what the \citetalias{LopezSanjuan15} method can be applied to in terms of typical galaxy number counts per field.

\section{Conclusions}
\label{conclusion}

In this paper we have presented observational applications of two different methods outlined by \citet{LopezSanjuan15} and \citet{Robertson10}, in which galaxy bias is measured from the enhanced scatter of a counts-in-cells distribution of galaxies due to `cosmic' variance. We have applied these methods to photometric candidates at redshift $z\sim2$ from the \emph{Hubble Space Telescope}'s Brightest of Reionizing Galaxies (BoRG) pure-parallel survey, the first application of these methods to such a survey. The BoRG survey consists of 141 independent fields covering $\sim$700 arcmin$^2$ with multiple broad-band optical and near-IR filters in the wavelength range 0.3 -- 1.7 $\mu$m on WFC3.

Using \texttt{SExtractor} we identified resolved sources in the $H_{160}$ near-IR filter and obtained multi-band photometry over this wavelength range. Considering only sources brighter than $m_{AB}=24.5$, we applied a colour selection ($Y_{098}-H_{160} > 1.5$ for BoRG[z8], $Y_{105}-H_{160} > 1.2$ for BoRG[z9]) to preselect for $z\sim2$ galaxies with the rest-frame 4000 \AA{} break spectral feature falling between these bands.

We obtain photometric redshifts for sources passing this cut using \texttt{BPZ}, identifying $N\sim1000$ candidates across the 141 fields in the survey with a `best redshift' in the redshift range $1.5 < z < 2.5$, yielding a mean redshift $\bar{z}=2.02$ and standard deviation $\sigma_z=0.25$, after applying quality-of-fit restrictions.

We applied the two methods to the resultant set of field counts, yielding galaxy bias values of $b \approx 2.87 \pm 0.43$ and $b \approx 3.63 \pm 0.57$ for the  \citetalias{LopezSanjuan15} and \citetalias{Robertson10} methods respectively for the complete sample, broadly consistent with measurements of the bias at this redshift conducted with correlation functions \citep[\textit{e.g.}][]{Wake11}. Dividing our sample into $N\sim400$ early and $N\sim600$ late type candidates based on SED template fits used in the photometric redshift determinations, we then examined the dependence of the bias on spectral type. This yielded galaxy bias values of $b_\textit{early} \approx 4.06 \pm 0.67$ and $b_\textit{late} \approx 2.98 \pm 0.98$ with the \citetalias{Robertson10} method, in line with correlation function measurements of pBzKs and sBzKs at $z\sim2$ \citep[\textit{e.g.}][]{Blanc08}. We found that this divided sample yielded number counts that were simply too low for an adequate log-normal fit to be obtained, preventing the proper implementation of the \citetalias{LopezSanjuan15} method to this part of the analysis.

Comparing the performance of the two methods, we conclude that the \citetalias{LopezSanjuan15} method works best on samples with higher number counts per field, where the effects of Poisson shot noise are suppressed compared to the effects of cosmic variance. The simpler \citetalias{Robertson10} method is more robust for applications where number counts per field are low. We conclude that our sample lies somewhat near the crossover where both methods can reasonably be applied, albeit yielding marginally different results.

The broad agreement of the galaxy bias estimates we have obtained with the \citetalias{LopezSanjuan15} and \citetalias{Robertson10} methods with existing measurements conducted with correlation functions demonstrates observationally for the first time the viability of this method as a tool to measure the bias of a galaxy population at high redshift from cosmic variance with space-based pure-parallel surveys. This is of particular interest with the upcoming launch of the \emph{James Webb Space Telescope} (\emph{JWST}). Predictions for \emph{JWST} based on the \citet{Mason15} luminosity function model indicate that a multi-band NIRCam pure-parallel survey similar to BoRG could yield $\sim$50 $z\sim6$ and $\sim$7 $z\sim8$ candidates per field per 3 hours of observation time to magnitude limit of $m_\text{AB} \lesssim 28$. Counts of this order would enable the measurement of the galaxy bias with the counts-in-cell approach used in this paper, achieving comparable uncertainties to those reported here. As such, these measurements would be a valuable tool for constraining models of the formation and evolution of galaxies during the Epoch of Reionization.

\section*{Acknowledgements}

Based on observations made with the NASA/ESA Hubble Space Telescope, which is operated by the Association of Universities for Research in Astronomy, Inc., under NASA contract NAS 5-26555. These observations are associated with programs 14701, 13767, 12905, 12572, and 11700. This research was supported by the Australian Research Council Centre of Excellence for All Sky Astrophysics in 3 Dimensions (ASTRO 3D), through project number CE170100013. AJC acknowledges support from an Australian Government Research Training Program (RTP) Scholarship. RCL acknowledges support from an Australian Research Council Discovery Early Career Researcher Award (DE180101240). We would also like to thank the referee for their constructive and insightful comments.



\bibliographystyle{mnras}
\bibliography{borg_bias} 

\begin{thebibliography}{}
\makeatletter
\relax
\def\mn@urlcharsother{\let\do\@makeother \do\$\do\&\do\#\do\^\do\_\do\%\do\~}
\def\mn@doi{\begingroup\mn@urlcharsother \@ifnextchar [ {\mn@doi@}
  {\mn@doi@[]}}
\def\mn@doi@[#1]#2{\def\@tempa{#1}\ifx\@tempa\@empty \href
  {http://dx.doi.org/#2} {doi:#2}\else \href {http://dx.doi.org/#2} {#1}\fi
  \endgroup}
\def\mn@eprint#1#2{\mn@eprint@#1:#2::\@nil}
\def\mn@eprint@arXiv#1{\href {http://arxiv.org/abs/#1} {{\tt arXiv:#1}}}
\def\mn@eprint@dblp#1{\href {http://dblp.uni-trier.de/rec/bibtex/#1.xml}
  {dblp:#1}}
\def\mn@eprint@#1:#2:#3:#4\@nil{\def\@tempa {#1}\def\@tempb {#2}\def\@tempc
  {#3}\ifx \@tempc \@empty \let \@tempc \@tempb \let \@tempb \@tempa \fi \ifx
  \@tempb \@empty \def\@tempb {arXiv}\fi \@ifundefined
  {mn@eprint@\@tempb}{\@tempb:\@tempc}{\expandafter \expandafter \csname
  mn@eprint@\@tempb\endcsname \expandafter{\@tempc}}}

\bibitem[\protect\citeauthoryear{{Adelberger}, {Steidel}, {Giavalisco},
  {Dickinson}, {Pettini}  \& {Kellogg}}{{Adelberger}
  et~al.}{1998}]{Adelberger98}
{Adelberger} K.~L.,  {Steidel} C.~C.,  {Giavalisco} M.,  {Dickinson} M.,
  {Pettini} M.,   {Kellogg} M.,  1998, \mn@doi [\apj] {10.1086/306162}, \href
  {http://adsabs.harvard.edu/abs/1998ApJ...505...18A} {505, 18}

\bibitem[\protect\citeauthoryear{{Adelberger}, {Steidel}, {Pettini}, {Shapley},
  {Reddy}  \& {Erb}}{{Adelberger} et~al.}{2005}]{Adelberger05}
{Adelberger} K.~L.,  {Steidel} C.~C.,  {Pettini} M.,  {Shapley} A.~E.,  {Reddy}
  N.~A.,   {Erb} D.~K.,  2005, \mn@doi [\apj] {10.1086/426580}, \href
  {http://adsabs.harvard.edu/abs/2005ApJ...619..697A} {619, 697}

\bibitem[\protect\citeauthoryear{{Andreani}, {Cristiani}, {Lucchin},
  {Matarrese}  \& {Moscardini}}{{Andreani} et~al.}{1994}]{Andreani94}
{Andreani} P.,  {Cristiani} S.,  {Lucchin} F.,  {Matarrese} S.,   {Moscardini}
  L.,  1994, \mn@doi [\apj] {10.1086/174422}, \href
  {http://adsabs.harvard.edu/abs/1994ApJ...430..458A} {430, 458}

\bibitem[\protect\citeauthoryear{{Barone-Nugent} et~al.,}{{Barone-Nugent}
  et~al.}{2014}]{BaroneNugent14}
{Barone-Nugent} R.~L.,  et~al., 2014, \mn@doi [\apj]
  {10.1088/0004-637X/793/1/17}, \href
  {http://adsabs.harvard.edu/abs/2014ApJ...793...17B} {793, 17}

\bibitem[\protect\citeauthoryear{{Ben{\'{\i}}tez}}{{Ben{\'{\i}}tez}}{2000}]{Benitez00}
{Ben{\'{\i}}tez} N.,  2000, \mn@doi [\apj] {10.1086/308947}, \href
  {http://adsabs.harvard.edu/abs/2000ApJ...536..571B} {536, 571}

\bibitem[\protect\citeauthoryear{{Ben{\'{\i}}tez} et~al.,}{{Ben{\'{\i}}tez}
  et~al.}{2004}]{Benitez04}
{Ben{\'{\i}}tez} N.,  et~al., 2004, \mn@doi [\apjs] {10.1086/380120}, \href
  {http://adsabs.harvard.edu/abs/2004ApJS..150....1B} {150, 1}

\bibitem[\protect\citeauthoryear{{Bertin} \& {Arnouts}}{{Bertin} \&
  {Arnouts}}{1996}]{SExtractor}
{Bertin} E.,  {Arnouts} S.,  1996, \mn@doi [\aaps] {10.1051/aas:1996164}, \href
  {http://adsabs.harvard.edu/abs/1996A%26AS..117..393B} {117, 393}

\bibitem[\protect\citeauthoryear{{Blanc} et~al.,}{{Blanc}
  et~al.}{2008}]{Blanc08}
{Blanc} G.~A.,  et~al., 2008, \mn@doi [\apj] {10.1086/588018}, \href
  {http://adsabs.harvard.edu/abs/2008ApJ...681.1099B} {681, 1099}

\bibitem[\protect\citeauthoryear{{Bradley} et~al.,}{{Bradley}
  et~al.}{2012}]{Bradley12}
{Bradley} L.~D.,  et~al., 2012, \mn@doi [\apj] {10.1088/0004-637X/760/2/108},
  \href {http://adsabs.harvard.edu/abs/2012ApJ...760..108B} {760, 108}

\bibitem[\protect\citeauthoryear{{Budav{\'a}ri} et~al.,}{{Budav{\'a}ri}
  et~al.}{2003}]{Budavari03}
{Budav{\'a}ri} T.,  et~al., 2003, \mn@doi [\apj] {10.1086/377168}, \href
  {http://adsabs.harvard.edu/abs/2003ApJ...595...59B} {595, 59}

\bibitem[\protect\citeauthoryear{{Calvi} et~al.,}{{Calvi}
  et~al.}{2016}]{Calvi16}
{Calvi} V.,  et~al., 2016, \mn@doi [\apj] {10.3847/0004-637X/817/2/120}, \href
  {http://adsabs.harvard.edu/abs/2016ApJ...817..120C} {817, 120}

\bibitem[\protect\citeauthoryear{{Cameron}, {Carollo}, {Oesch}, {Bouwens},
  {Illingworth}, {Trenti}, {Labb{\'e}}  \& {Magee}}{{Cameron}
  et~al.}{2011}]{CameronE11}
{Cameron} E.,  {Carollo} C.~M.,  {Oesch} P.~A.,  {Bouwens} R.~J.,
  {Illingworth} G.~D.,  {Trenti} M.,  {Labb{\'e}} I.,   {Magee} D.,  2011,
  \mn@doi [\apj] {10.1088/0004-637X/743/2/146}, \href
  {http://adsabs.harvard.edu/abs/2011ApJ...743..146C} {743, 146}

\bibitem[\protect\citeauthoryear{{Casertano} et~al.,}{{Casertano}
  et~al.}{2000}]{Casertano00}
{Casertano} S.,  et~al., 2000, \mn@doi [\aj] {10.1086/316851}, \href
  {http://adsabs.harvard.edu/abs/2000AJ....120.2747C} {120, 2747}

\bibitem[\protect\citeauthoryear{{Coe}, {Ben{\'{\i}}tez}, {S{\'a}nchez}, {Jee},
  {Bouwens}  \& {Ford}}{{Coe} et~al.}{2006}]{Coe06}
{Coe} D.,  {Ben{\'{\i}}tez} N.,  {S{\'a}nchez} S.~F.,  {Jee} M.,  {Bouwens} R.,
    {Ford} H.,  2006, \mn@doi [\aj] {10.1086/505530}, \href
  {http://adsabs.harvard.edu/abs/2006AJ....132..926C} {132, 926}

\bibitem[\protect\citeauthoryear{{Coil}, {Mendez}, {Eisenstein}  \&
  {Moustakas}}{{Coil} et~al.}{2017}]{Coil17}
{Coil} A.~L.,  {Mendez} A.~J.,  {Eisenstein} D.~J.,   {Moustakas} J.,  2017,
  \mn@doi [\apj] {10.3847/1538-4357/aa63ec}, \href
  {http://adsabs.harvard.edu/abs/2017ApJ...838...87C} {838, 87}

\bibitem[\protect\citeauthoryear{{Colless} et~al.,}{{Colless}
  et~al.}{2001}]{Colless01}
{Colless} M.,  et~al., 2001, \mn@doi [\mnras]
  {10.1046/j.1365-8711.2001.04902.x}, \href
  {http://adsabs.harvard.edu/abs/2001MNRAS.328.1039C} {328, 1039}

\bibitem[\protect\citeauthoryear{{Daddi}, {Cimatti}, {Renzini}, {Fontana},
  {Mignoli}, {Pozzetti}, {Tozzi}  \& {Zamorani}}{{Daddi}
  et~al.}{2004}]{Daddi04}
{Daddi} E.,  {Cimatti} A.,  {Renzini} A.,  {Fontana} A.,  {Mignoli} M.,
  {Pozzetti} L.,  {Tozzi} P.,   {Zamorani} G.,  2004, \mn@doi [\apj]
  {10.1086/425569}, \href {http://adsabs.harvard.edu/abs/2004ApJ...617..746D}
  {617, 746}

\bibitem[\protect\citeauthoryear{{Dodelson}}{{Dodelson}}{2004}]{Dodelson04}
{Dodelson} S.,  2004, \mn@doi [\prd] {10.1103/PhysRevD.70.023009}, \href
  {http://adsabs.harvard.edu/abs/2004PhRvD..70b3009D} {70, 023009}

\bibitem[\protect\citeauthoryear{{Efstathiou}}{{Efstathiou}}{1995}]{Efstathiou95}
{Efstathiou} G.,  1995, \mn@doi [\mnras] {10.1093/mnras/276.4.1425}, \href
  {http://adsabs.harvard.edu/abs/1995MNRAS.276.1425E} {276, 1425}

\bibitem[\protect\citeauthoryear{{Efstathiou}, {Kaiser}, {Saunders},
  {Lawrence}, {Rowan-Robinson}, {Ellis}  \& {Frenk}}{{Efstathiou}
  et~al.}{1990}]{Efstathiou90}
{Efstathiou} G.,  {Kaiser} N.,  {Saunders} W.,  {Lawrence} A.,
  {Rowan-Robinson} M.,  {Ellis} R.~S.,   {Frenk} C.~S.,  1990, \mnras, \href
  {http://adsabs.harvard.edu/abs/1990MNRAS.247P..10E} {247, 10P}

\bibitem[\protect\citeauthoryear{{Harikane} et~al.,}{{Harikane}
  et~al.}{2018}]{Harikane18}
{Harikane} Y.,  et~al., 2018, \mn@doi [\pasj] {10.1093/pasj/psx097}, \href
  {http://adsabs.harvard.edu/abs/2018PASJ...70S..11H} {70, S11}

\bibitem[\protect\citeauthoryear{{Hayashi}, {Shimasaku}, {Motohara}, {Yoshida},
  {Okamura}  \& {Kashikawa}}{{Hayashi} et~al.}{2007}]{Hayashi07}
{Hayashi} M.,  {Shimasaku} K.,  {Motohara} K.,  {Yoshida} M.,  {Okamura} S.,
  {Kashikawa} N.,  2007, \mn@doi [\apj] {10.1086/513068}, \href
  {http://adsabs.harvard.edu/abs/2007ApJ...660...72H} {660, 72}

\bibitem[\protect\citeauthoryear{{Ishikawa}, {Kashikawa}, {Toshikawa},
  {Tanaka}, {Hamana}, {Niino}, {Ichikawa}  \& {Uchiyama}}{{Ishikawa}
  et~al.}{2017}]{Ishikawa17}
{Ishikawa} S.,  {Kashikawa} N.,  {Toshikawa} J.,  {Tanaka} M.,  {Hamana} T.,
  {Niino} Y.,  {Ichikawa} K.,   {Uchiyama} H.,  2017, \mn@doi [\apj]
  {10.3847/1538-4357/aa6d64}, \href
  {http://adsabs.harvard.edu/abs/2017ApJ...841....8I} {841, 8}

\bibitem[\protect\citeauthoryear{{Kent} \& {Gunn}}{{Kent} \&
  {Gunn}}{1982}]{Kent&Gunn82}
{Kent} S.~M.,  {Gunn} J.~E.,  1982, \mn@doi [\aj] {10.1086/113178}, \href
  {http://adsabs.harvard.edu/abs/1982AJ.....87..945K} {87, 945}

\bibitem[\protect\citeauthoryear{{Koekemoer}, {Fruchter}, {Hook}  \&
  {Hack}}{{Koekemoer} et~al.}{2003}]{Koekemoer03}
{Koekemoer} A.~M.,  {Fruchter} A.~S.,  {Hook} R.~N.,   {Hack} W.,  2003, in
  {Arribas} S.,  {Koekemoer} A.,   {Whitmore} B.,  eds, HST Calibration
  Workshop : Hubble after the Installation of the ACS and the NICMOS Cooling
  System. p.~337

\bibitem[\protect\citeauthoryear{{Lee}, {Giavalisco}, {Gnedin}, {Somerville},
  {Ferguson}, {Dickinson}  \& {Ouchi}}{{Lee} et~al.}{2006}]{Lee06}
{Lee} K.-S.,  {Giavalisco} M.,  {Gnedin} O.~Y.,  {Somerville} R.~S.,
  {Ferguson} H.~C.,  {Dickinson} M.,   {Ouchi} M.,  2006, \mn@doi [\apj]
  {10.1086/500387}, \href {http://adsabs.harvard.edu/abs/2006ApJ...642...63L}
  {642, 63}

\bibitem[\protect\citeauthoryear{{Li}, {Kauffmann}, {Jing}, {White},
  {B{\"o}rner}  \& {Cheng}}{{Li} et~al.}{2006}]{Li06}
{Li} C.,  {Kauffmann} G.,  {Jing} Y.~P.,  {White} S.~D.~M.,  {B{\"o}rner} G.,
  {Cheng} F.~Z.,  2006, \mn@doi [\mnras] {10.1111/j.1365-2966.2006.10066.x},
  \href {http://adsabs.harvard.edu/abs/2006MNRAS.368...21L} {368, 21}

\bibitem[\protect\citeauthoryear{{Lin} et~al.,}{{Lin} et~al.}{2012}]{Lin12}
{Lin} L.,  et~al., 2012, \mn@doi [\apj] {10.1088/0004-637X/756/1/71}, \href
  {http://adsabs.harvard.edu/abs/2012ApJ...756...71L} {756, 71}

\bibitem[\protect\citeauthoryear{{L{\'o}pez-Sanjuan}
  et~al.,}{{L{\'o}pez-Sanjuan} et~al.}{2014}]{LopezSanjuan14}
{L{\'o}pez-Sanjuan} C.,  et~al., 2014, \mn@doi [\aap]
  {10.1051/0004-6361/201322474}, \href
  {http://adsabs.harvard.edu/abs/2014A%26A...564A.127L} {564, A127}

\bibitem[\protect\citeauthoryear{{L{\'o}pez-Sanjuan}
  et~al.,}{{L{\'o}pez-Sanjuan} et~al.}{2015}]{LopezSanjuan15}
{L{\'o}pez-Sanjuan} C.,  et~al., 2015, \mn@doi [\aap]
  {10.1051/0004-6361/201526731}, \href
  {http://adsabs.harvard.edu/abs/2015A%26A...582A..16L} {582, A16}

\bibitem[\protect\citeauthoryear{{Madgwick} et~al.,}{{Madgwick}
  et~al.}{2003}]{Madgwick03}
{Madgwick} D.~S.,  et~al., 2003, \mn@doi [\mnras]
  {10.1046/j.1365-8711.2003.06861.x}, \href
  {http://adsabs.harvard.edu/abs/2003MNRAS.344..847M} {344, 847}

\bibitem[\protect\citeauthoryear{{Mason}, {Trenti}  \& {Treu}}{{Mason}
  et~al.}{2015}]{Mason15}
{Mason} C.~A.,  {Trenti} M.,   {Treu} T.,  2015, \mn@doi [\apj]
  {10.1088/0004-637X/813/1/21}, \href
  {http://adsabs.harvard.edu/abs/2015ApJ...813...21M} {813, 21}

\bibitem[\protect\citeauthoryear{{McCracken} et~al.,}{{McCracken}
  et~al.}{2010}]{McCracken10}
{McCracken} H.~J.,  et~al., 2010, \mn@doi [\apj] {10.1088/0004-637X/708/1/202},
  \href {http://adsabs.harvard.edu/abs/2010ApJ...708..202M} {708, 202}

\bibitem[\protect\citeauthoryear{{Meneux} et~al.,}{{Meneux}
  et~al.}{2009}]{Meneux09}
{Meneux} B.,  et~al., 2009, \mn@doi [\aap] {10.1051/0004-6361/200912314}, \href
  {http://adsabs.harvard.edu/abs/2009A%26A...505..463M} {505, 463}

\bibitem[\protect\citeauthoryear{{Millington} \& {Peach}}{{Millington} \&
  {Peach}}{1986}]{Millington86}
{Millington} S.~J.~C.,  {Peach} J.~V.,  1986, \mn@doi [\mnras]
  {10.1093/mnras/221.1.15}, \href
  {http://adsabs.harvard.edu/abs/1986MNRAS.221...15M} {221, 15}

\bibitem[\protect\citeauthoryear{{Mo} \& {White}}{{Mo} \&
  {White}}{1996}]{Mo&White96}
{Mo} H.~J.,  {White} S.~D.~M.,  1996, \mn@doi [\mnras]
  {10.1093/mnras/282.2.347}, \href
  {http://adsabs.harvard.edu/abs/1996MNRAS.282..347M} {282, 347}

\bibitem[\protect\citeauthoryear{{Mo} \& {White}}{{Mo} \&
  {White}}{2002}]{Mo&White02}
{Mo} H.~J.,  {White} S.~D.~M.,  2002, \mn@doi [\mnras]
  {10.1046/j.1365-8711.2002.05723.x}, \href
  {http://adsabs.harvard.edu/abs/2002MNRAS.336..112M} {336, 112}

\bibitem[\protect\citeauthoryear{{Mostek}, {Coil}, {Cooper}, {Davis}, {Newman}
  \& {Weiner}}{{Mostek} et~al.}{2013}]{Mostek13}
{Mostek} N.,  {Coil} A.~L.,  {Cooper} M.,  {Davis} M.,  {Newman} J.~A.,
  {Weiner} B.~J.,  2013, \mn@doi [\apj] {10.1088/0004-637X/767/1/89}, \href
  {http://adsabs.harvard.edu/abs/2013ApJ...767...89M} {767, 89}

\bibitem[\protect\citeauthoryear{{Norberg} et~al.,}{{Norberg}
  et~al.}{2001}]{Norberg01}
{Norberg} P.,  et~al., 2001, \mn@doi [\mnras]
  {10.1046/j.1365-8711.2001.04839.x}, \href
  {http://adsabs.harvard.edu/abs/2001MNRAS.328...64N} {328, 64}

\bibitem[\protect\citeauthoryear{{Norberg} et~al.,}{{Norberg}
  et~al.}{2002}]{Norberg02}
{Norberg} P.,  et~al., 2002, \mn@doi [\mnras]
  {10.1046/j.1365-8711.2002.05348.x}, \href
  {http://adsabs.harvard.edu/abs/2002MNRAS.332..827N} {332, 827}

\bibitem[\protect\citeauthoryear{{Oke} \& {Gunn}}{{Oke} \&
  {Gunn}}{1983}]{Oke&Gunn83}
{Oke} J.~B.,  {Gunn} J.~E.,  1983, \mn@doi [\apj] {10.1086/160817}, \href
  {http://adsabs.harvard.edu/abs/1983ApJ...266..713O} {266, 713}

\bibitem[\protect\citeauthoryear{{Ouchi} et~al.,}{{Ouchi}
  et~al.}{2004}]{Ouchi04}
{Ouchi} M.,  et~al., 2004, \mn@doi [\apj] {10.1086/422208}, \href
  {http://adsabs.harvard.edu/abs/2004ApJ...611..685O} {611, 685}

\bibitem[\protect\citeauthoryear{{Ouchi} et~al.,}{{Ouchi}
  et~al.}{2005}]{Ouchi05}
{Ouchi} M.,  et~al., 2005, \mn@doi [\apjl] {10.1086/499519}, \href
  {http://adsabs.harvard.edu/abs/2005ApJ...635L.117O} {635, L117}

\bibitem[\protect\citeauthoryear{{Peebles}}{{Peebles}}{1980}]{Peebles80}
{Peebles} P.~J.~E.,  1980, {The large-scale structure of the universe}

\bibitem[\protect\citeauthoryear{{Planck Collaboration} et~al.,}{{Planck
  Collaboration} et~al.}{2016}]{Planck16}
{Planck Collaboration} et~al., 2016, \mn@doi [A&A]
  {10.1051/0004-6361/201527101}, 594, A1

\bibitem[\protect\citeauthoryear{{Robertson}}{{Robertson}}{2010}]{Robertson10}
{Robertson} B.~E.,  2010, \mn@doi [\apjl] {10.1088/2041-8205/716/2/L229}, \href
  {http://adsabs.harvard.edu/abs/2010ApJ...716L.229R} {716, L229}

\bibitem[\protect\citeauthoryear{{Rood}}{{Rood}}{1979}]{Rood79}
{Rood} H.~J.,  1979, \mn@doi [\apj] {10.1086/157328}, \href
  {http://adsabs.harvard.edu/abs/1979ApJ...232..699R} {232, 699}

\bibitem[\protect\citeauthoryear{{Sato}, {Sawicki}  \& {Arcila-Osejo}}{{Sato}
  et~al.}{2014}]{Sato14}
{Sato} T.,  {Sawicki} M.,   {Arcila-Osejo} L.,  2014, \mn@doi [\mnras]
  {10.1093/mnras/stu1356}, \href
  {http://adsabs.harvard.edu/abs/2014MNRAS.443.2661S} {443, 2661}

\bibitem[\protect\citeauthoryear{{Schlafly} \& {Finkbeiner}}{{Schlafly} \&
  {Finkbeiner}}{2011}]{Schlafly11}
{Schlafly} E.~F.,  {Finkbeiner} D.~P.,  2011, \mn@doi [\apj]
  {10.1088/0004-637X/737/2/103}, \href
  {http://adsabs.harvard.edu/abs/2011ApJ...737..103S} {737, 103}

\bibitem[\protect\citeauthoryear{{Schmidt} et~al.,}{{Schmidt}
  et~al.}{2014}]{Schmidt14}
{Schmidt} K.~B.,  et~al., 2014, \mn@doi [\apj] {10.1088/0004-637X/786/1/57},
  \href {http://adsabs.harvard.edu/abs/2014ApJ...786...57S} {786, 57}

\bibitem[\protect\citeauthoryear{{Skibba} et~al.,}{{Skibba}
  et~al.}{2014}]{Skibba14}
{Skibba} R.~A.,  et~al., 2014, \mn@doi [\apj] {10.1088/0004-637X/784/2/128},
  \href {http://adsabs.harvard.edu/abs/2014ApJ...784..128S} {784, 128}

\bibitem[\protect\citeauthoryear{{Steidel}, {Giavalisco}, {Pettini},
  {Dickinson}  \& {Adelberger}}{{Steidel} et~al.}{1996}]{Steidel96}
{Steidel} C.~C.,  {Giavalisco} M.,  {Pettini} M.,  {Dickinson} M.,
  {Adelberger} K.~L.,  1996, \mn@doi [\apjl] {10.1086/310029}, \href
  {http://adsabs.harvard.edu/abs/1996ApJ...462L..17S} {462, L17}

\bibitem[\protect\citeauthoryear{{Trenti} \& {Stiavelli}}{{Trenti} \&
  {Stiavelli}}{2008}]{Trenti08}
{Trenti} M.,  {Stiavelli} M.,  2008, \mn@doi [\apj] {10.1086/528674}, \href
  {http://adsabs.harvard.edu/abs/2008ApJ...676..767T} {676, 767}

\bibitem[\protect\citeauthoryear{{Trenti} et~al.,}{{Trenti}
  et~al.}{2011}]{Trenti11}
{Trenti} M.,  et~al., 2011, \mn@doi [\apjl] {10.1088/2041-8205/727/2/L39},
  \href {http://adsabs.harvard.edu/abs/2011ApJ...727L..39T} {727, L39}

\bibitem[\protect\citeauthoryear{{Trenti} et~al.,}{{Trenti}
  et~al.}{2012}]{Trenti12}
{Trenti} M.,  et~al., 2012, \mn@doi [\apj] {10.1088/0004-637X/746/1/55}, \href
  {http://adsabs.harvard.edu/abs/2012ApJ...746...55T} {746, 55}

\bibitem[\protect\citeauthoryear{{Tyson}, {Valdes}  \& {Wenk}}{{Tyson}
  et~al.}{1990}]{Tyson90}
{Tyson} J.~A.,  {Valdes} F.,   {Wenk} R.~A.,  1990, \mn@doi [\apjl]
  {10.1086/185636}, \href {http://adsabs.harvard.edu/abs/1990ApJ...349L...1T}
  {349, L1}

\bibitem[\protect\citeauthoryear{{Viironen} et~al.,}{{Viironen}
  et~al.}{2018}]{Viironen18}
{Viironen} K.,  et~al., 2018, \mn@doi [\aap] {10.1051/0004-6361/201731797},
  \href {http://adsabs.harvard.edu/abs/2018A%26A...614A.129V} {614, A129}

\bibitem[\protect\citeauthoryear{{Wake} et~al.,}{{Wake} et~al.}{2011}]{Wake11}
{Wake} D.~A.,  et~al., 2011, \mn@doi [\apj] {10.1088/0004-637X/728/1/46}, \href
  {http://adsabs.harvard.edu/abs/2011ApJ...728...46W} {728, 46}

\bibitem[\protect\citeauthoryear{{Wang}, {Yang}, {Mo}  \& {van den
  Bosch}}{{Wang} et~al.}{2007}]{Wang07}
{Wang} Y.,  {Yang} X.,  {Mo} H.~J.,   {van den Bosch} F.~C.,  2007, \mn@doi
  [\apj] {10.1086/519245}, \href
  {http://adsabs.harvard.edu/abs/2007ApJ...664..608W} {664, 608}

\bibitem[\protect\citeauthoryear{{York} et~al.,}{{York} et~al.}{2000}]{York00}
{York} D.~G.,  et~al., 2000, \mn@doi [\aj] {10.1086/301513}, \href
  {http://adsabs.harvard.edu/abs/2000AJ....120.1579Y} {120, 1579}

\bibitem[\protect\citeauthoryear{{Zehavi} et~al.,}{{Zehavi}
  et~al.}{2005}]{Zehavi05}
{Zehavi} I.,  et~al., 2005, \mn@doi [\apj] {10.1086/431891}, \href
  {http://adsabs.harvard.edu/abs/2005ApJ...630....1Z} {630, 1}

\bibitem[\protect\citeauthoryear{{Zehavi} et~al.,}{{Zehavi}
  et~al.}{2011}]{Zehavi11}
{Zehavi} I.,  et~al., 2011, \mn@doi [\apj] {10.1088/0004-637X/736/1/59}, \href
  {http://adsabs.harvard.edu/abs/2011ApJ...736...59Z} {736, 59}

\bibitem[\protect\citeauthoryear{{de la Torre} et~al.,}{{de la Torre}
  et~al.}{2011}]{delaTorre11}
{de la Torre} S.,  et~al., 2011, \mn@doi [\mnras]
  {10.1111/j.1365-2966.2010.17939.x}, \href
  {http://adsabs.harvard.edu/abs/2011MNRAS.412..825D} {412, 825}

\makeatother
\end{thebibliography}



\appendix

\section{Field Counts} 
\label{field_counts}

A full list of $z\sim2$ candidate counts from the fields analysed is given in tables \ref{tab:z8_counts} and \ref{tab:z9_counts}.

\begin{table}
	\centering
	\caption{Table of $z\sim2$ candidate counts from the BoRG[z8] data set. Raw candidate counts are given in the second column. Areas are given in arcmin$^{2}$. The final column shows count rescaled to correspond to an area of 4.41 arcmin$^{2}$}
	\label{tab:z8_counts}
	\begin{tabular}{lccc} 
		\hline
		Field name & $z\sim2$ & Area & Scaled count\\
		 & & arcmin$^{2}$ & \\
		\hline		
		0110-0224 & 10 & 13.81 & 3.19\\
		0214+1255 & 1 & 4.42 & 1.00\\
		0228-4102 & 4 & 4.43 & 3.98\\
		0240-1857 & 5 & 4.45 & 4.96\\
		0436-5259 & 4 & 4.33 & 4.07\\
		0439-5317 & 5 & 4.28 & 5.15\\
		0440-5244 & 5 & 4.34 & 5.08\\
		0456-2203 & 13 & 4.40 & 13.02\\
		0540-6409 & 7 & 3.95 & 7.82\\
		0553-6405 & 4 & 4.00 & 4.41\\
		0624-6432 & 5 & 4.44 & 4.96\\
		0624-6440 & 5 & 4.39 & 5.02\\
		0637-7518 & 13 & 6.33 & 9.06\\
		0751+2917 & 8 & 4.52 & 7.80\\
		0756+3043 & 4 & 4.19 & 4.21\\
		0808+3946 & 7 & 4.30 & 7.18\\
		0819+4911 & 6 & 4.52 & 5.85\\
		0820+2332 & 7 & 4.46 & 6.91\\
		0835+2456 & 10 & 4.43 & 9.95\\
		0846+7654 & 11 & 4.41 & 11.00\\
		0906+0255 & 8 & 4.39 & 8.03\\
		0909+0002 & 6 & 4.40 & 6.02\\
		0914+2822 & 12 & 4.40 & 12.02\\
		0922+4505 & 3 & 4.38 & 3.02\\
		0926+4000 & 4 & 4.45 & 3.96\\
		0926+4426 & 7 & 4.49 & 6.88\\
		0951+3304 & 8 & 4.41 & 8.00\\
		0952+5304 & 4 & 4.42 & 3.99\\
		1010+3001 & 11 & 4.54 & 10.68\\
		1014-0423 & 1 & 4.48 & 0.98\\
		1031+3804 & 3 & 4.48 & 2.95\\
		1031+5052 & 8 & 5.55 & 6.36\\
		1033+5051 & 6 & 5.50 & 4.81\\
		1051+3359 & 12 & 4.26 & 12.43\\
		1059+0519 & 9 & 4.43 & 8.96\\
		1103-2330 & 9 & 4.37 & 9.09\\
		1111+5545 & 6 & 4.31 & 6.15\\
		1118-1858 & 3 & 4.23 & 3.13\\
		1119+4026 & 7 & 4.46 & 6.92\\
		1131+3114 & 6 & 4.41 & 5.99\\
		1152+5441 & 5 & 4.40 & 5.02\\
		1153+0056 & 4 & 4.44 & 3.97\\
		1209+4543 & 6 & 4.42 & 5.98\\
		1230+0750 & 6 & 4.24 & 6.24\\
		1242+5716 & 10 & 4.29 & 10.29\\
		1245+3356 & 10 & 4.43 & 9.95\\
		1301+0000 & 1 & 4.44 & 0.99\\
		1337+0028 & 6 & 4.47 & 5.92\\
		1341+4123 & 7 & 4.36 & 7.07\\
		1358+4326 & 14 & 4.49 & 13.74\\
		1358+4334 & 8 & 4.32 & 8.16\\
		1408+5503 & 4 & 4.32 & 4.08\\
		1416+1638 & 17 & 4.38 & 17.12\\
		1429-0331 & 8 & 4.35 & 8.10\\
		1437+5043 & 9 & 6.53 & 6.08\\
		1459+7146 & 12 & 4.32 & 12.26\\
		1510+1115 & 14 & 4.43 & 13.92\\
		1524+0954 & 6 & 4.32 & 6.12\\
		1555+1108 & 7 & 4.31 & 7.16\\
		1632+3733 & 2 & 4.37 & 2.02\\
		\hline
	\end{tabular}
\end{table}

\begin{table}
	\centering
	\contcaption{Table of $z\sim2$ candidate counts from the BoRG[z8] data set. Raw candidate counts are given in the second column. Areas are given in arcmin$^{2}$. The final column shows count rescaled to correspond to an area of 4.41 arcmin$^{2}$}
	\label{tab:z8_counts_cont}
	\begin{tabular}{lccc} 
		\hline
		Field name & $z\sim2$ & Area & Scaled count\\
		& & arcmin$^{2}$ & \\
		\hline
		1632+3737 & 7 & 4.37 & 7.06\\
		2057-4412 & 8 & 4.34 & 8.13\\
		2132+1004 & 3 & 4.46 & 2.96\\
		2132-1202 & 8 & 4.37 & 8.08\\
		2155-4411 & 10 & 4.49 & 9.82\\
		2203+1851 & 8 & 4.60 & 7.66\\
		2313-2243 & 3 & 5.59 & 2.37\\
		2345+0054 & 2 & 4.48 & 1.97\\
		2351-4332 & 18 & 4.30 & 18.47\\
		\hline
	\end{tabular}
\end{table}
		
\begin{table}
	\centering
	\caption{Table of $z\sim2$ candidate counts from the BoRG[z9] data set. Raw candidate counts are given in the second column. Areas are given in arcmin$^{2}$. The final column shows count rescaled to correspond to an area of 4.41 arcmin$^{2}$}
	\label{tab:z9_counts}
	\begin{tabular}{lccc} 
		\hline
		Field name & $z\sim2$ & Area & Scaled count\\
		 & & arcmin$^{2}$ & \\
		\hline
		0116+1425 & 6 & 4.29 & 6.17\\
		0119-3411 & 0 & 4.07 & 0.00\\
		0132-7326 & 10 & 4.31 & 10.23\\
		0235-0357 & 4 & 5.93 & 2.97\\
		0314-6712 & 13 & 4.64 & 12.36\\
		0337-0507 & 5 & 4.50 & 4.89\\
		0554-6005 & 11 & 4.16 & 11.67\\
		0751+2917 & 9 & 4.46 & 8.91\\
		0807+3606 & 3 & 4.46 & 2.96\\
		0851+4240 & 9 & 4.37 & 9.07\\
		0853+0310 & 3 & 4.45 & 2.97\\
		0925+1360 & 7 & 4.44 & 6.95\\
		0925+3439 & 4 & 4.43 & 3.98\\
		0949+5759 & 5 & 4.26 & 5.18\\
		0953+5150 & 8 & 4.42 & 7.98\\
		0953+5153 & 8 & 4.58 & 7.71\\
		0953+5157 & 5 & 4.43 & 4.98\\
		0955+4528 & 2 & 4.48 & 1.97\\
		0956+2848 & 5 & 4.43 & 4.98\\
		1017-2052 & 9 & 4.37 & 9.08\\
		1018+0544 & 8 & 4.40 & 8.02\\
		1048+1518 & 10 & 4.40 & 10.02\\
		1103+2913 & 7 & 4.41 & 7.00\\
		1104+2813 & 7 & 4.37 & 7.06\\
		1106+2855 & 4 & 4.43 & 3.98\\
		1106+2925 & 7 & 4.46 & 6.93\\
		1106+3508 & 6 & 4.39 & 6.02\\
		1115+2548 & 7 & 4.49 & 6.87\\
		1136+0747 & 6 & 4.42 & 5.98\\
		1142+2640 & 4 & 4.50 & 3.92\\
		1142+2647 & 8 & 4.42 & 7.98\\	
		1142+3020 & 5 & 4.45 & 4.96\\
		1143+3019 & 9 & 4.40 & 9.03\\
		1149+2202 & 1 & 4.40 & 1.00\\
		1152+3402 & 3 & 4.31 & 3.07\\
		1152+5434 & 13 & 5.69 & 10.08\\
		1154+4639 & 12 & 4.30 & 12.32\\
		1160+0015 & 10 & 4.44 & 9.94\\
		1209+4543 & 6 & 4.40 & 6.01\\
		1218+3008 & 7 & 4.31 & 7.16\\
		\hline
	\end{tabular}
\end{table}

\begin{table}
	\centering
	\contcaption{Table of $z\sim2$ candidate counts from the BoRG[z9] data set. Raw candidate counts are given in the second column. Areas are given in arcmin$^{2}$. The final column shows count rescaled to correspond to an area of 4.41 arcmin$^{2}$}
	\label{tab:z9_counts_cont}
	\begin{tabular}{lccc} 
		\hline
		Field name & $z\sim2$ & Area & Scaled count\\
		& & arcmin$^{2}$ & \\
		\hline
		1259+4128 & 14 & 6.06 & 10.19\\
		1334+3131 & 3 & 4.50 & 2.94\\
		1410+2623 & 9 & 4.40 & 9.02\\
		1413+0918 & 2 & 4.35 & 2.03\\
		1431+0259 & 0 & 4.08 & 0.00\\
		1437-0150 & 7 & 4.35 & 7.09\\
		1438-0142 & 8 & 4.44 & 7.95\\
		1442-0212 & 8 & 4.41 & 7.99\\
		1503+3645 & 5 & 4.32 & 5.11\\
		1519-0746 & 2 & 4.44 & 1.99\\
		1520-2501 & 1 & 4.29 & 1.03\\
		1524+0956 & 2 & 4.34 & 2.03\\
		1525+0955 & 12 & 4.29 & 12.34\\
		1525+0960 & 5 & 4.28 & 5.16\\
		1536+1410 & 6 & 4.45 & 5.94\\
		1558+0812 & 18 & 4.34 & 18.30\\
		1607+1332 & 6 & 4.43 & 5.97\\
		1614+4856 & 3 & 4.38 & 3.02\\
		1619+2541 & 4 & 4.33 & 4.08\\
		1632+3736 & 7 & 4.65 & 6.64\\
		1659+3732 & 6 & 4.29 & 6.17\\
		1708+4237 & 6 & 4.33 & 6.11\\
		1715+0455 & 4 & 4.25 & 4.15\\
		1715+0502 & 3 & 4.25 & 3.11\\
		1738+1839 & 7 & 4.22 & 7.32\\
		2008-6610 & 8 & 4.24 & 8.32\\
		2057-1423 & 7 & 4.29 & 7.20\\
		2134-0708 & 11 & 4.47 & 10.85\\
		2140+0241 & 6 & 4.44 & 5.96\\
		2228-0955 & 4 & 4.42 & 3.99\\
		2229-0945 & 23 & 4.43 & 22.88\\
		2253-1411 & 10 & 4.45 & 9.91\\
		2312-1423 & 5 & 4.42 & 4.99\\
		2323+0059 & 6 & 4.10 & 6.45\\
		\hline
	\end{tabular}
\end{table}

\section{Primary Targets of the Most Overdense Fields} 
\label{overdense}

The coordinates of each pointing in a pure-parallel survey such as the BoRG survey are determined by those of a primary observation. These primary observations will generally be drawn from a large variety of different programs resulting in a set of of pointings distributed randomly across the sky. This randomness is important for our clustering measurement as we require a sample of fields that is not systematically biased toward regions of high density.

However, specific primary observations targeting regions of high density have the potential to yield pointings with high number counts that could be considered contaminants to this analysis. An example of this would be a redshift $z \sim 2$ quasar. Such an object could be expected to reside in a high density environment and an observation taken within $\sim$6 arcmin of this could bias our sample. 

We have summarised the primary observations of fields with scaled counts of 12 or higher in table \ref{tab:most_overdense} to ensure our analysis is not significantly biased in this way. This cut-off corresponds to a probability of $P(n\geq12) \approx 0.05$ according to the associated Poisson distribution. Two of these fields were imaged in parallel to a primary targeting either an exo-planetary system (ExoPl) or a galactic star. Galactic targets will have no impact in biasing this survey. Similarly, low redshift galaxies, such as the redshift $z=0.167$ galaxy targeted in this set of fields, are not an issue. The majority had primary programs targeting quasars (QSO). However, all of these targets had redshifts of $z<1$ (much lower than our redshift range), with the exception of one field (2351-4332). In this case the quasar is at a redshift of $z=2.885$, sufficiently higher than our range of interest to avoid any impact. The remaining target was an absorption line system (AbLS) outside our redshift range of interest ($z=1.278$). Thus, we concluded that none of the most overdense fields was physically associated to the primary target of the HST observations. 

\begin{table}
	\centering
	\caption{Summary of the primary targets of the most overdense fields from the survey. The \textit{Obj. Type} column describes the type of object targeted by the primary observation (QSO is a quasar, ExoPl is an exo-planetary system, Star a galactic star, Gal a galaxy, and AbLS an absorption line system) while column $z$ gives its redshift. The field name and its density of counts per 4.41 arcmin$^{2}$ are given for reference.}
	\label{tab:most_overdense}
	\begin{tabular}{lccc} 
		\hline
		Field name & Scaled count & \multicolumn{2}{c}{Primary target} \\
		&  & Obj. Type & $z$ \\
		\hline
		2229-0945 & 22.88 & QSO & 0.798\\
		2351-4332 & 18.47 & QSO & 2.885\\
		1558+0812 & 18.30 & QSO & 0.517\\
		1416+1638 & 17.12 & QSO & 0.744\\
		1510+1115 & 13.92 & QSO & 0.285\\
		1358+4326 & 13.74 & ExoPl & galactic\\
		0456-2203 & 13.02 & QSO & 0.533\\
		1051+3359 & 12.43 & Gal & 0.167\\
		0314-6712 & 12.36 & Star & galactic\\
		1525+0955 & 12.34 & AbLS & 1.278\\
		1154+4639 & 12.32 & QSO & 0.643\\
		1459+7146 & 12.26 & QSO & 0.905\\
		0914+2822 & 12.02 & QSO & 0.735\\
		\hline
	\end{tabular}
\end{table}


\bsp	
\label{lastpage}
\end{document}